\documentclass[useAMS,usenatbib,usegraphicx]{mn2e}

\newcommand{\heii}{HeII $\lambda$4686}

\newcommand{\ha}{H$\alpha$}
\newcommand{\hb}{H$\beta$}

\newcommand{\bb}{Bowen blend}
\newcommand{\kms}{km s$^{-1}$}
\newcommand{\msun}{M$_{\odot}$}
\newcommand{\aap}{A\&A}
\newcommand{\mnras}{MNRAS}
\newcommand{\aj}{AJ}
\newcommand{\apj}{ApJ}
\newcommand{\apjl}{ApJL}
\newcommand{\pasp}{PASP}
\newcommand{\araa}{ARA\&A}             

\title[Optical Spectroscopy of GX 339-4 - Paper I]{Optical Spectroscopy of GX 339-4 - Paper I: Orbital Modulation}
\author[M. Buxton and S. Vennes]{M. Buxton$^{1}$\thanks{Now at Astronomy Department, Yale University.  E-mail:  buxton@astro.yale.edu} and S. Vennes$^{2}$\\
$^{1}$Research School of Astronomy \& Astrophysics, Mount Stromlo Observatory, Cotter Road, Weston Creek, ACT, 2611, Australia\\
$^{2}$Department of Mathematics, John Dedman Building, Australian National University, Acton, ACT, 0200, Australia}
\begin{document}

\date{}

\pagerange{\pageref{firstpage}--\pageref{lastpage}} \pubyear{2002}

\maketitle

\label{firstpage}

\begin{abstract}
Optical spectroscopic observations of GX 339-4 were carried out between 1998 May (X-ray high state) and 1999 May (X-ray low state) over 7 epochs.  The equivalent width of {\ha} increased during the high state then decreased during the low state.  The equivalent width of {\heii} decreased over both states.  The full-width half-maximum of {\ha}, {\hb}, {\heii} and the {\bb} increased from 1998 to 1999 indicating that the emission line regions moved closer to the compact object.  {\hb} shows a redshifted absorption feature at $\lambda$4880 at all epochs.  This line remains unidentified.  Analysis of individual spectra from 1998 May 28-31 show modulation of the radial velocities, equivalent width and V/R ratios of {\ha} on the 14.86 hour orbital period.  The equivalent width of {\heii} also varies on the orbital period.  This is the first time since the study of \citet{cal92} that spectroscopic data has confirmed the orbital period.  The semi-amplitude of the {\ha} radial velocities is $K_{1}$ = 14 {\kms}.  Hence the mass function = 2 x 10$^{-4}$ {\msun}.  

\end{abstract}


\section{Introduction}

GX 339-4 was discovered by $OSO-7$ in 1973 \citep{mar73}.  The secondary star has a magnitude B $>$ 21 \citep*{hut81, ilo81}.  Consequently the secondary has never been directly observed and its spectral type never identified.  

An orbital period of 14.86 hours (0.61916$\pm$0.0027 days) was derived by \citet[][hereafter CCHT92]{cal92} from $R$- and $I$-band photometry taken during an X-ray off state.  Two sets of spectra obtained in the high state by \citet*{cow87} were folded on this period by CCHT92.  The first data set did not show any modulation but the second set did, giving a semi-amplitude $K_1$ = 78 $\pm$ 13 \kms\ and systemic velocity $\gamma$ = -62 $\pm$ \kms.  The mass of the compact object was estimated to be $M_1 \le$ 1 {\msun} based on these measurements and assuming the binary inclination $i \le 70^o$.  This estimate of $M_1$ is well below the theoretical maximum mass of a neutron star \citep[$\sim$ 3.2 \msun,][]{chi76, fri87}.  GX 339-4 has been classified as a black hole X-ray binary based purely on X-ray characteristics and behaviour which are very similar to another black-hole candidate Cyg X-1.   If GX 339-4 is not black hole then this places the method of black-hole classification using X-ray data into serious question.

CCHT92 proposed that the optical modulation is due to eclipses of an X-ray heated companion and/or accretion disk as it was seen in both the off and high state data.  This would require $i \ge$ 60$^o$ which has not been supported by other optical or X-ray observations.  

\textit{The orbital period has not been confirmed by either optical spectroscopy or photometry since the CCHT92 study.}  This means that the orbital period is wrong, that the orbital period is only detectable under certain physical conditions in the binary or that the data quality of subsequent studies has not been sufficient.  

The main aim of this study was to determine the orbital period of GX 339-4 from emission lines in optical spectra.  This method is notoriously difficult.  As the emission lines originate from the accretion disk information such as the radial velocities are more difficult to extract since their amplitudes are much smaller and they may be contaminated by other flux sources.  Since we have no spectral signature of the secondary star we are left with no other choice but to use disk emission lines to derive the orbital period.

In this paper we present the results of analysis conducted on the emission lines in GX 339-4.  We discuss the spectral changes seen over the epochs and how these changes relate to the X-ray behaviour over the same period.  In addition we present results of analysis aimed at detecting the orbital period from disk emission line variations.

\section[]{Observations and Data Reduction}

A total of 51 low dispersion and 59 high dispersion spectra were obtained between 1998 May and 1999 May.  An observation log is given in Table \ref{tab:obslog}.  Observations were performed using the Double-Beam Spectrograph (red and blue arms) on the Australian National University 2.3m telescope and the Royal Greenwich Observatory spectrograph on the Anglo-Australian Telescope.

\begin{table*}
\begin{centering}
\caption{Observation log of GX 339-4 optical spectroscopy.}
\label{tab:obslog}
\begin{tabular}{l|l|l|r|r|r|l}
\hline
 Date & Telescope & Grating & Resolution & Exposure & Wavelength Range & No. of\\
 & & (l/mm) & (\AA) & (sec) & (\AA) & Spectra\\
\hline
1998 May 26,27 & ANU 2.3m & 300/316 & 4.4 & 900,1800,2700,3600 & 3300-9500 & 20\\
1998 May 28-31 & ANU 2.3m & 1200 & 2.2 & 3600 & 4600-5300, 6100-7050 &  25 \\
1998 Aug 20-23 & ANU 2.3m & 1200 & 2.2 & 2000 & 4400-5100, 6200-7150 & 17 \\
1999 Mar 11    & ANU 2.3m & 600 & 4.4 & 900 & 3150-8800 & 5 \\
1999 Mar 24-27 & ANU 2.3m & 300/316 & 4.4-8.4 & 900 & 3300-9450 & 20 \\
1999 Apr 13    & AAT 3.9m & 1200 & 1.5 & 1800 & 6480-7290 & 3 \\
1999 Apr 17, 18 & ANU 2.3m & 1200 & 2.2 & 1800 & 4250-4970, 6050-7000 & 14 \\
1999 May 11    & ANU 2.3m & 300/316 & 4.4 & 1800 & 3300-9450 & 6 \\
\hline
\end{tabular}
\end{centering}
\end{table*}

\begin{figure*}
\begin{centering}
\includegraphics[width=12cm,height=14cm]{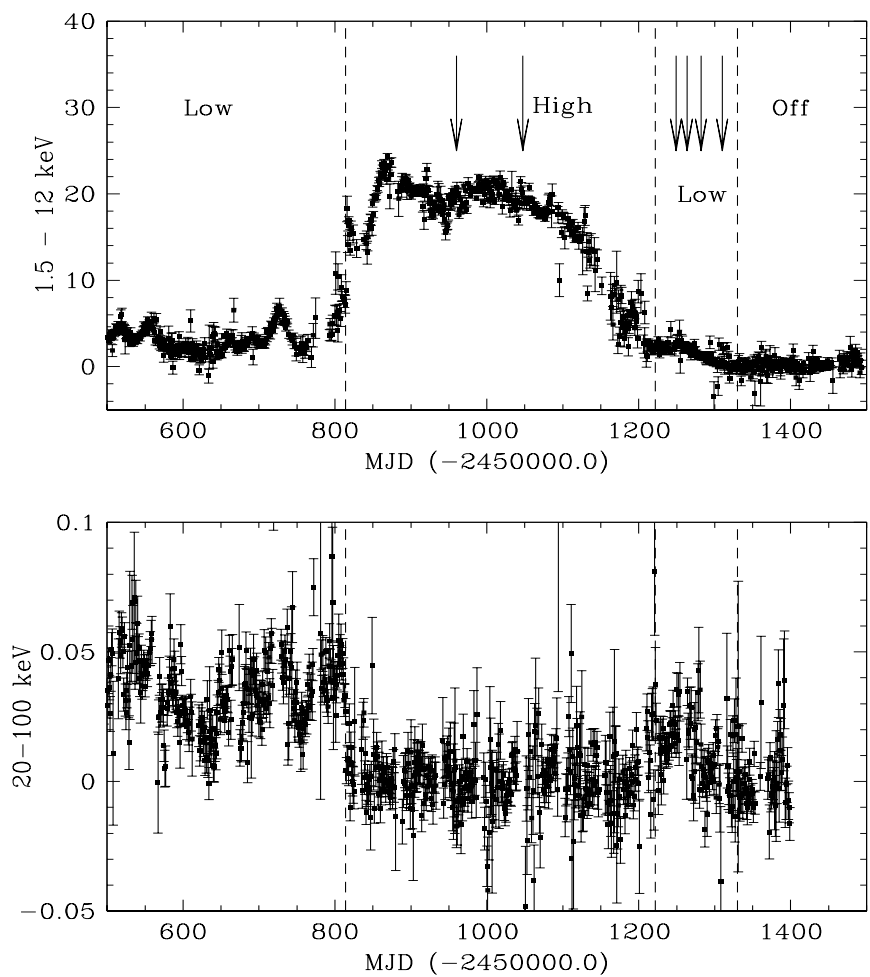}
   \caption{$Top$: The ASM day-averaged count rate (counts/sec) of GX 339-4 in 1.5 - 12 keV band (1 Crab $\sim$ 75 counts/sec).  Quick-look results provided by the ASM/RXTE team.  $Bottom$: BATSE photon fluxes in 20 - 100 keV band.  Arrows indicate epochs of optical observations.  `Low', `High' and `Off' refer to the X-ray states.  Transition times (indicated by the vertical dashed lines) and X-ray state classifications were taken from \citet{bel99} and \citet{kon00}.}
\label{fig:gx-xray}
\end{centering}
\end{figure*}

Figure \ref{fig:gx-xray} shows the epochs of observations with respect to the soft (1.5-12 keV) and hard (20-100 keV) X-ray observations.  Our observations of GX 339-4 were obtained during high and low X-ray states. 

Data reduction was performed using IRAF.  Each spectrum was corrected for electronic bias and pixel-to-pixel variations.  The background was fitted using a low-order polynomial.  Wavelength calibration was performed via FeAr (blue) and NeAr (red) arc spectra from which line positions were fitted with a fourth-order polynomial.  The largest rms error on the wavelength calibration for blue spectra was 0.01 {\AA} ($\sim$ 1 \kms) in high resolution spectra and 0.8 {\AA} ($\sim$ 50 \kms) in low resolution spectra.  In red spectra the largest rms error was 0.01 {\AA} ($\sim$ 0.5 \kms) in high resolution spectra and 0.4 {\AA} ($\sim$ 18 \kms) in low resolution spectra.  Telluric lines were removed using a smooth-spectrum standard star applying the method outlined in \citet{bes99}.  For nights which had clear observing conditions spectra were flux calibrated using the standard stars Feige 110, LTT 7379, HD 49798 and HD 127493.

\section{Spectra in High and Low X-ray States}
\label{sec:shls}

\begin{figure*}
\begin{centering}
\includegraphics[width=8cm,height=9cm]{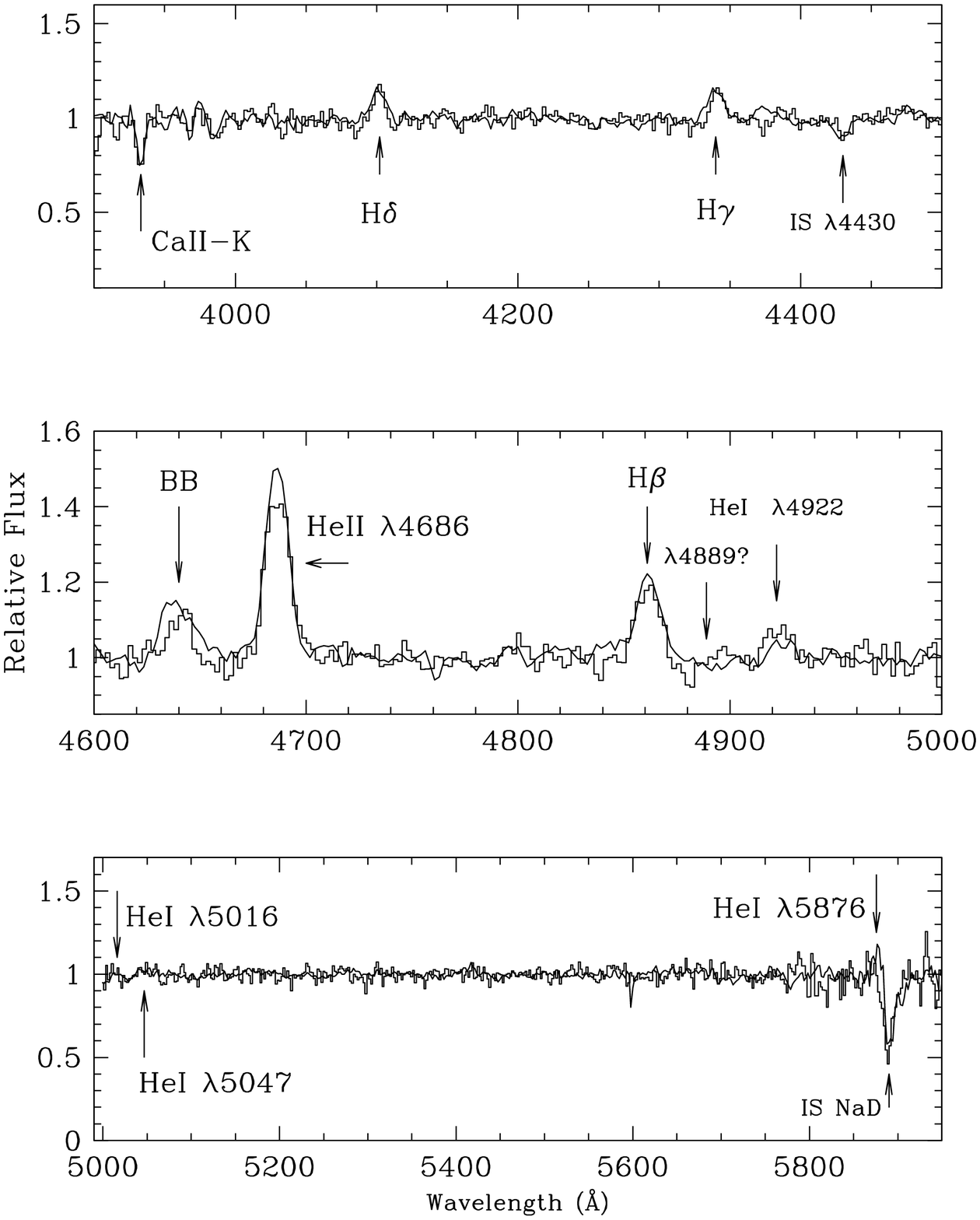}
\includegraphics[width=8cm,height=9cm]{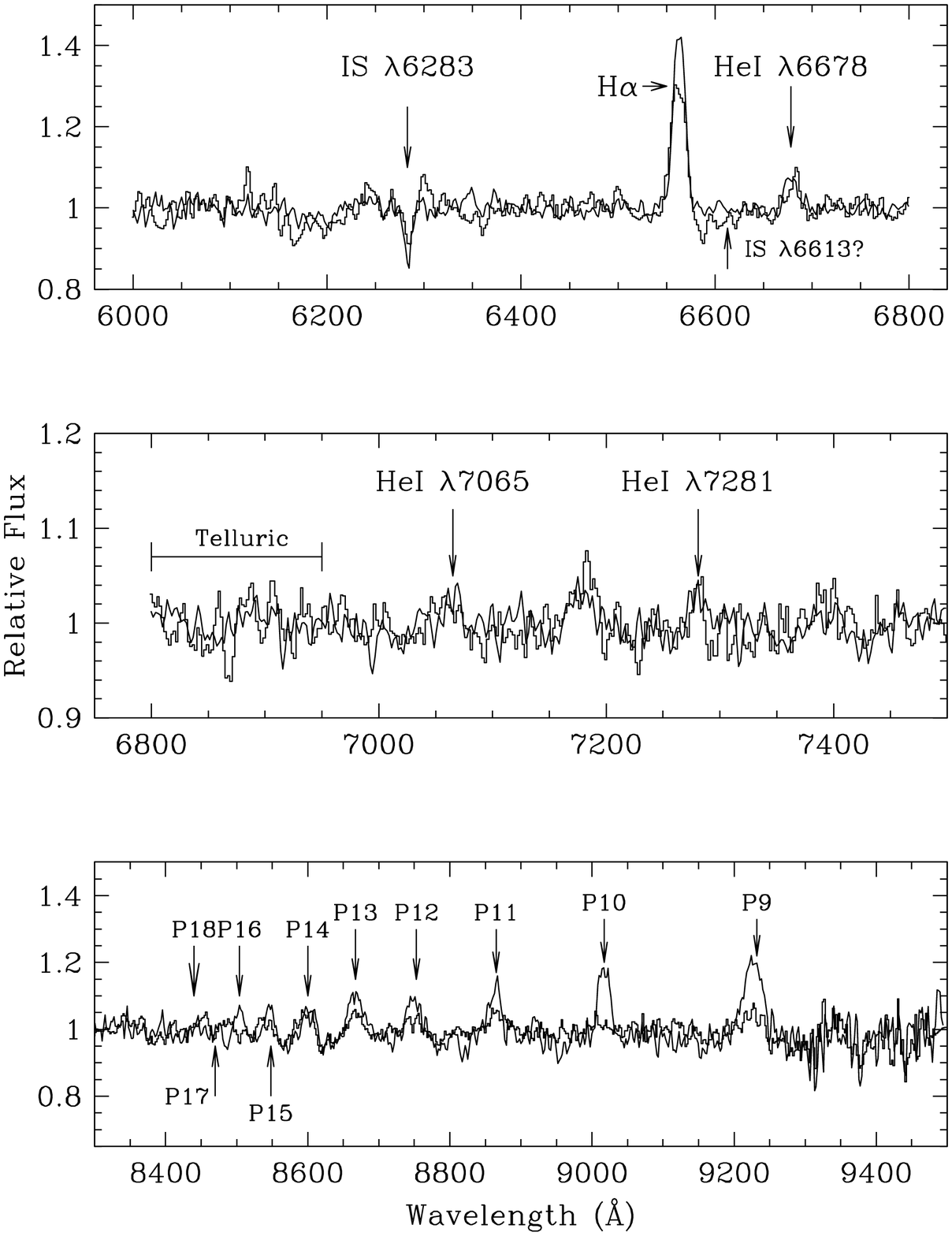}
   \caption{Left three panels are blue spectra and right three panels red spectra of GX 339-4 in the high (solid line) and low (histogram) X-ray states.  Visible lines are indicated.  Low resolution data only are shown here to present full wavelength coverage.  `BB' is the Bowen blend.  'Telluric' indicates region of strongest telluric lines which may not have been removed perfectly.}
\label{fig:gx-bluehilo}
\end{centering}
\end{figure*}

Figure \ref{fig:gx-bluehilo} shows combined, normalised blue and red spectra from the high and low X-ray states.  Only low dispersion spectra are shown here to present full wavelength coverage.  Spectra were combined with weighting on individual spectra according to their signal-to-noise ratio (SNR).  

Many emission lines are present in the high state.  The strongest lines are H$\alpha$ and HeII $\lambda$4686.  Other emission lines visible include the H$\gamma$, H$\beta$, the Bowen blend ($\sim \lambda$4640 \AA), HeII $\lambda$5411, HeI $\lambda\lambda$4922, 5875 (blended with NaD), 6678 and 7281 and the Paschen series.  

In the low state most of the lines are still visible.  The most dramatic flux decrease from the high state to the low state is seen in H$\delta$, the {\bb}, {\heii} and the lower members of the Paschen series. 

The absorption lines present are CaII-K ($\lambda$3934), the NaD $\lambda \lambda$5889, 5895 lines and the diffuse lines $\lambda \lambda$4430 and 6283.  These are of interstellar origin.  

\begin{figure}
\begin{centering}
\includegraphics[width=7cm,height=8cm]{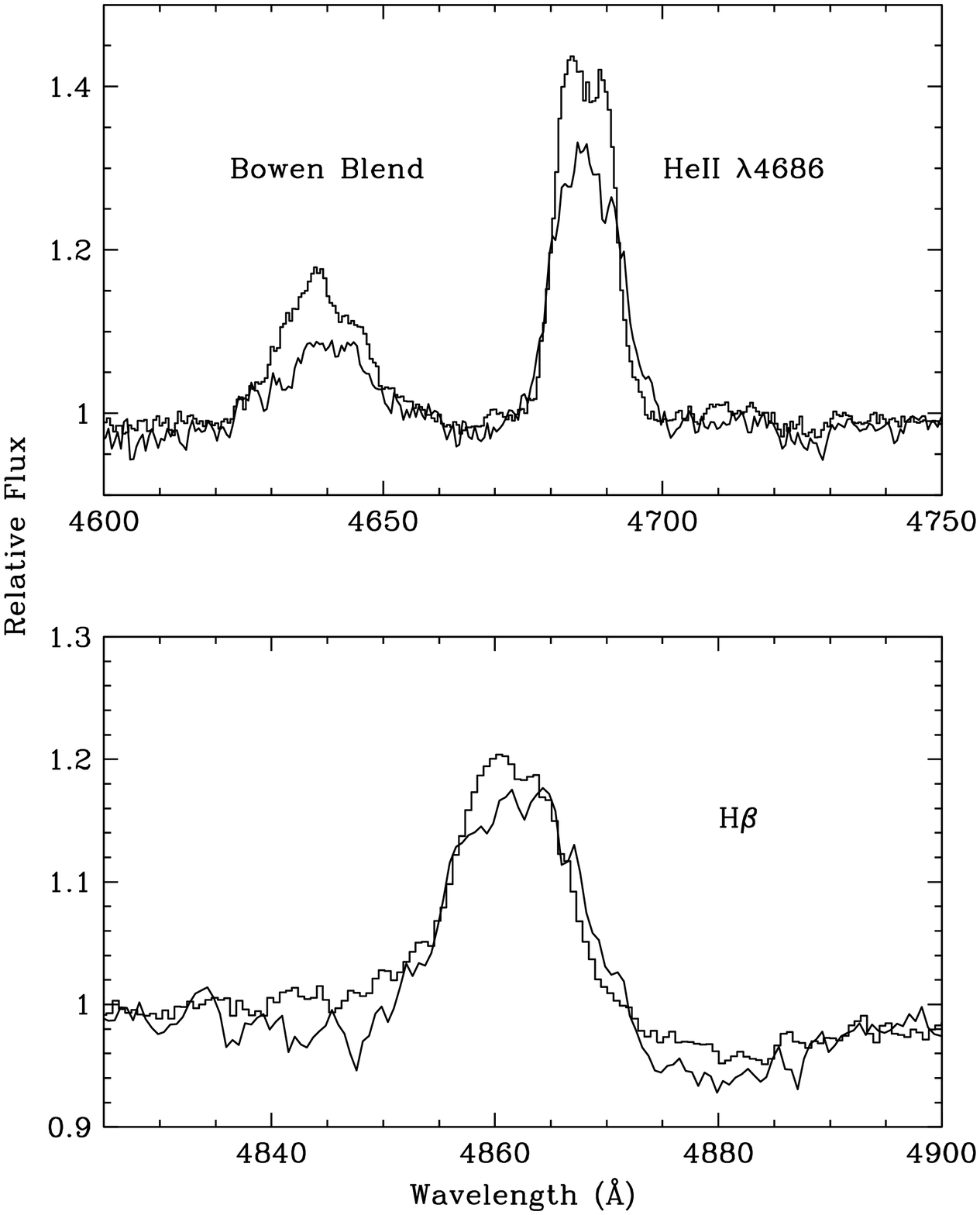}
\includegraphics[width=7cm,height=8cm]{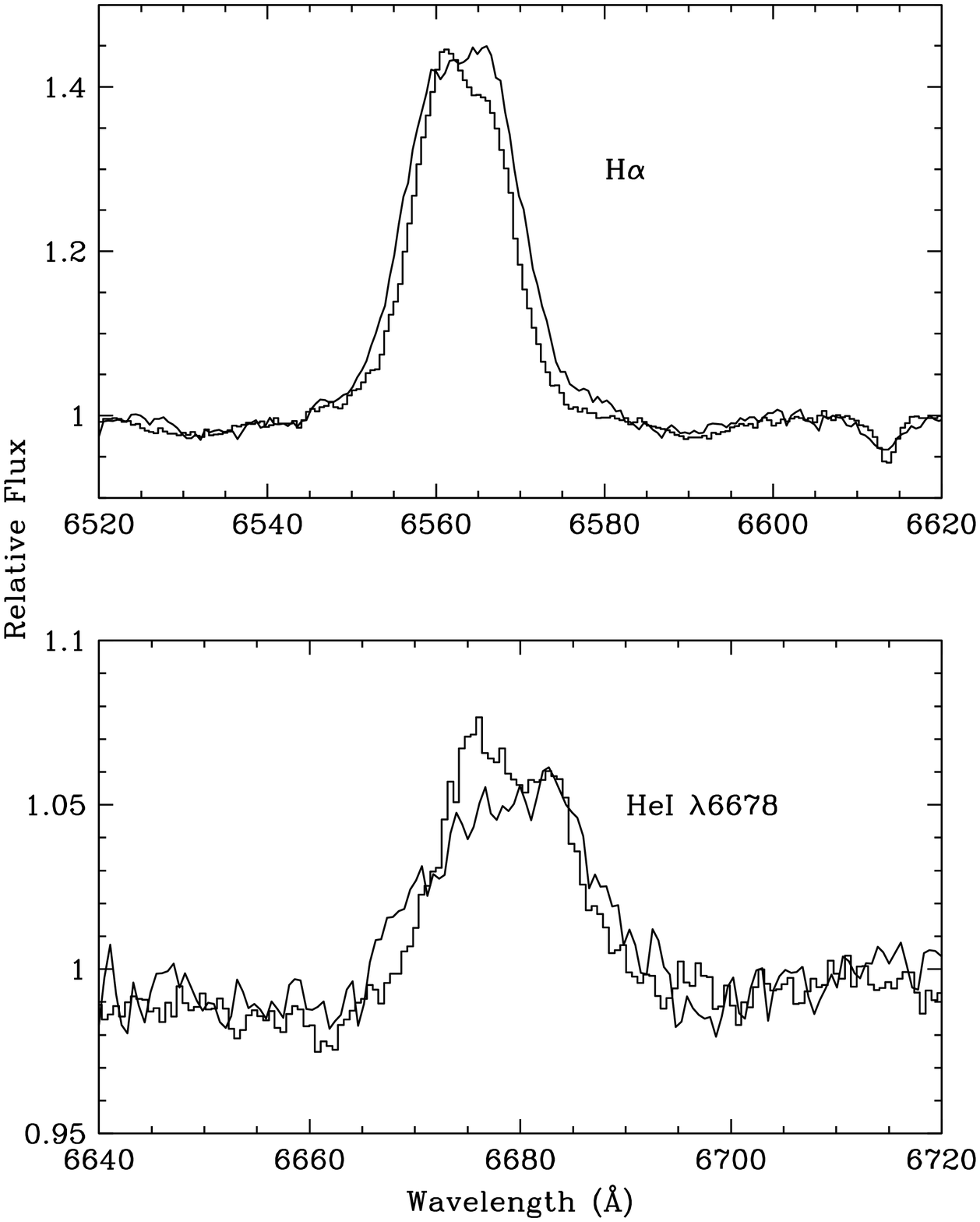}
   \caption{Combined high and low resolution spectra from high (histogram) and low states (line) expanded around the most dominant emission lines.  }
\label{fig:comb_fwhm_hilob}
\end{centering}
\end{figure}

Figure \ref{fig:comb_fwhm_hilob} shows an expanded view of the dominant emission lines in combined spectra over the high and low states.  The combined spectra in this case include both high and low resolution data with weighting on the SNR of individual spectra.

An increase in line width is most obvious in {\ha} but less so in other lines.  The red wing of {\heii} and {\hb} have slightly increased.  The Balmer and HeI lines are asymmetric on the blue side in the high state.  They become more symmetrical in the low state. 

The {\hb} emission line shows a redshifted absorption dip in both the high and low state data.  Further discussion on this feature is given in \S \ref{sec:hb_abs}.

The similarities in flux behaviour between the {\bb} and {\heii} suggest that both emission lines arise from the Bowen Flourescence Mechanism \citep*[BFM,][]{mcc75}.  Other emission lines expected from the BFM are OIII $\lambda\lambda$3133 and 3444.  Spectra are too noisy in this region to ascertain whether these lines are present.  Weaker emission from NIII $\lambda$4097-4103 is also expected but would be blended with H$\delta$.  This could explain why H$\delta$ shows the most flux decrease of the Balmer lines. 

The decrease in {\heii} and {\bb} emission is most likely due to fewer soft X-rays available for photoionisation.  There must be enough, however, to sufficiently excite the lower ionisation lines of HeI and the Balmer series.  Parts of the disk which were heated in the high state may be allowed to cool sufficiently in the low state for Balmer and HeI emission to occur closer to the compact object.  This would explain the increase in line widths (see also \S \ref{sec:elvoe}).  

In the next sections we will present quantitatively how the emission lines vary on long (days to months) and short (hours) timescales. 

The analysis of the spectra has been divided into two parts.  The first deals with combined spectra from which the distance, extinction, magnitudes and colours are measured.  Consideration is also given as to how the equivalent width (EW) and full-width half maximum (FWHM) of the emission lines vary in comparison to the soft X-ray behaviour between 1998 and 1999. 

The second part of the analysis is concerned with individual spectra.  The analysis aims to determine any orbital modulation which may be present in the properties of the emission lines.

In a companion paper (Paper II) we will explore further the line profile behaviour and how the variations may be related to various sources in the binary.

\section{Analysis of Combined Spectra}
\label{sec:gx_spec_analysis}

\subsection{Distance and Extinction}
\label{sec:dist_ext}

Previous measurements of the distance $D$ and extinction $E(B-V)$ to GX 339-4 are summarised in Table \ref{tab:other_dist}.  

\begin{table*}
\begin{center}
\caption{Previous measurements of distance and extinction of GX 339-4.}
\label{tab:other_dist}
\begin{tabular}{l|c|c|c|c}
\hline
Reference & Distance & EW of NaD & EW of CaII-K & $E(B-V)$\\
 & (kpc) & (\AA) & (\AA) & \\
\hline
\citet{dox79} & $\sim$ 4 & & & \\
\citet{gri79} & $\sim$ 8 & 2.2 $\pm$ 0.4 & & $\sim$ 1.2 \\
\citet{mau86} & 1.33 & & & \\
\citet{mak86} & 3.5 & & & \\
\citet{ilo86} & & & & 0.7 $\pm$ 0.1 \\
\citet{cow87} & 3.5-4 & & 1.3 & \\
\citet{cor87} & & 1.0 & & \\
\citet{pre91} & 1.3 & & & \\
\citet{zdz98} & 4 $\pm$ 1 & & & \\
\citet*{sor99} & & $\sim$ 3.7 & & \\
\citet*{sha01} & 5.6 & & &  \\
\hline
 \end{tabular}
\end{center}
\end{table*}

Spectra were combined for nights when the NaD and CaII-K lines were included in the spectral range.  Four or five measurements of the EWs of NaD and CaII-K were made from the combined spectra for each night.  Measurements were obtained by fitting single-peaked Gaussians to the line profiles using \textit{splot} in IRAF, varying the placement of the continuum.  They are listed in Table \ref{tab:gx_is_ew}.  Errors were derived from the scatter in the measurements.  

The grand average EW of the NaD lines is 4.2 $\pm$ 1.0 {\AA} and for CaII-K 1.6 $\pm$ 0.3 \AA.  The former result agrees with \citet{sor99} but is higher than \citet{cor87} and \citet{gri79}.  The latter agrees well with \citet{cow87}.  

Using the method described in \citet{all73} we find that $D \sim$ 4 $\pm$ 1 kpc.  For CaII-K we used the empirical relationship from \citet{bea53} giving $D \sim$ 4 $\pm$ 1 kpc.  This agrees with the result from the NaD lines.  Beals \& Oke also provide a relationship for the NaD lines.  In this case $D \sim$ 7 $\pm$ 2 kpc which is larger than the other results but just consistent within the errors.  Since the first two results agree well with one another and with previous studies we have adopted $D = 4 \pm 1$ kpc.  The other measurement should not be ignored, however, and indeed highlights the inconsistency which various methods of distance determination via interstellar lines may give.

$E(B-V)$ was calculated using the empirical relationship of \citet{bar90} and was found to be $\sim$ 1.1 $\pm$ 0.2.  This is consistent with \citet{gri79}.  Using the interstellar extinction relationship of \citet{sav79}, $A_V$ = 3.1$E(B-V) = 3.3 \pm 0.7$.  

\begin{table}
\begin{center}
  \caption{Measured equivalent widths of NaD and CaII-K interstellar absorption lines from GX 339-4 spectra. }
\label{tab:gx_is_ew} 
  \begin{tabular}{r|c|c}
\hline
Date & EW of NaD & EW of CaII-K \\
& (\AA) & (\AA)\\
\hline
1998 May 26 & 3.8 $\pm$ 0.4 & 1.6 $\pm$ 0.4 \\
May 27 & 5.5 $\pm$ 1.5 & 1.7 $\pm$ 0.4 \\
1999 Mar 11 & 4.0 $\pm$ 0.6 & - \\
Mar 24 & 3.8 $\pm$ 1.2 & - \\
Mar 25 & 4.3 $\pm$ 1.1 & 1.5 $\pm$ 0.3 \\
Mar 26 & 4.5 $\pm$ 0.8 & 1.6 $\pm$ 0.2 \\
May 11 & 3.7 $\pm$ 0.5 & - \\
\hline
Average & 4.2 $\pm$ 1.0 & 1.6 $\pm$ 0.3\\
\hline
 \end{tabular}  
\end{center}
\end{table}

\subsection{Continuum Variations}
\label{sec:gx_ctm_var}

\begin{figure}
\begin{centering}
\includegraphics[width=7cm,height=8cm]{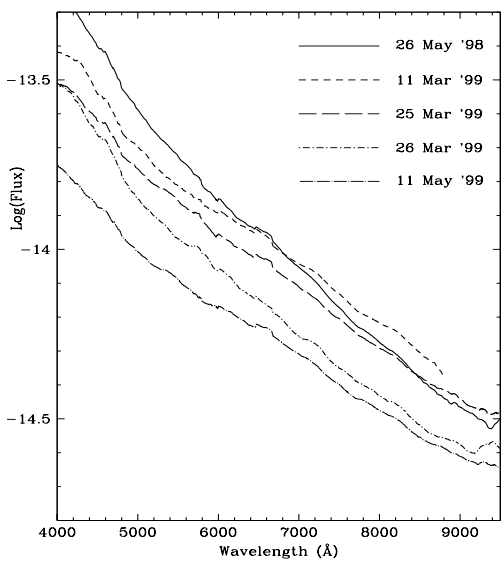}
  \caption{Flux-calibrated spectra over some observed epochs.  Flux units are ergs cm$^{-2}$ s$^{-1}$ \AA$^{-1}$.  Spectra have been heavily smoothed to accentuate the continuum behaviour.  The continuum decreases and reddens from 1998 to 1999 but the slope can vary greatly from night to night as seen with 1999 March 25 and 26 spectra.}
\label{fig:combdereddened}
\end{centering}
\end{figure}

Combined, flux-calibrated spectra are shown in Figure \ref{fig:combdereddened}.  Individual spectra were averaged with weighting according to their SNR.  The spectra were dereddened using $A_V = 3.3 \pm 0.7$ (\S \ref{sec:dist_ext}).  

The continuum is blue which is typical of XRBs during outburst.  The continuum flux decreased steadily over the year of observations. The continuum reddened as the binary went from the high state to low state (see also \S \ref{sec:colours}).  This is due to less optical flux originating from the accretion disk as fewer X-ray photons are available to be reprocessed by the outer disk.  From night to night, however, the slope varied dramatically as seen with spectra from 1999 March 25 and 26.  On these nights the slit width was 2 arcseconds and seeing conditions were noted as being 1.5 - 2 arcseconds.  Hence most of the flux from the system would have been recorded and changes in the continuum are probably real. 

\subsection{Optical Magnitudes and Colours}
\label{sec:colours}

$B$, $V$, $R$ and $I$ magnitudes and ($B-V$), ($V-R$) and ($V-\nolinebreak I$) colours were derived from the combined flux-calibrated, dereddened spectra using \begin{em}sbands\end{em} in IRAF.  The results are shown in Tables \ref{tab:spec_op_mags} and \ref{tab:spec_op_colours}.  Errors in the magnitudes were determined by finding the scatter between the sensitivity functions of the standard stars used in the flux calibration.  The errors in the colours were derived by measuring the difference in slopes between the sensitivity functions.

\begin{table*}
\begin{center}
\caption{Magnitudes of GX 339-4 measured from combined, dereddened, flux-calibrated spectra.}
\label{tab:spec_op_mags}
  \begin{tabular}{r|c|c|c|c}
\hline
Date & $B$ & $V$ & $R$ & $I$ \\
\hline
1998 May 26 & 12.89 $\pm$ 0.75 & 13.12 $\pm$ 0.75 & 12.85 $\pm$ 0.75 & 13.02 $\pm$ 0.75  \\
1999 Mar 11 & - & 13.35 $\pm$ 0.75 & 12.91 $\pm$ 0.75 & - \\
Mar 25 & 13.51 $\pm$ 0.75 & 13.49 $\pm$ 0.75 & 13.08 $\pm$ 0.75 & 13.05 $\pm$ 0.75  \\
Mar 26 & 13.57 $\pm$ 0.75 & 13.72 $\pm$ 0.75 & 13.37 $\pm$ 0.75 & 13.41 $\pm$ 0.75 \\
May 11 & 14.13 $\pm$ 0.75 & 14.07 $\pm$ 0.75 & 13.62 $\pm$ 0.75 & 13.51 $\pm$ 0.75  \\
\hline
 \end{tabular}
\end{center}
\end{table*}

\begin{table}
\begin{center}
\caption{Colours of GX 339-4 measured from combined, dereddened, flux-calibrated spectra.}
\label{tab:spec_op_colours}
\begin{tabular}{r|c|c|c}
\hline
Date & ($B-V$) & ($V-R$) & ($V-I$) \\
\hline
1998 May 26 & -0.23 $\pm$ 0.03 & 0.27 $\pm$ 0.05 & 0.10 $\pm$ 0.10 \\
1999 Mar 11 & - & 0.44 $\pm$ 0.05 & -  \\
Mar 25 & 0.02 $\pm$ 0.03 & 0.41 $\pm$ 0.05 & 0.44 $\pm$ 0.10 \\
Mar 26 & -0.15 $\pm$ 0.03 & 0.35 $\pm$ 0.05 & 0.31 $\pm$ 0.10  \\
May 11 & 0.06 $\pm$ 0.03 & 0.45 $\pm$ 0.05 & 0.56 $\pm$ 0.10 \\
\hline
 \end{tabular}
\end{center}
\end{table}

The trend from 1998 to 1999 suggests that the magnitudes have decreased.  The errors, however, are larger than the greatest difference between the magnitudes.  Hence we cannot say whether this trend is real.   

\begin{figure}
\begin{centering}
\includegraphics[width=7cm,height=8cm]{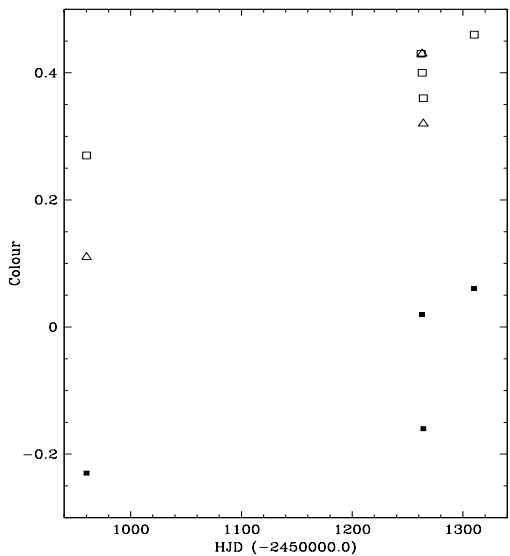}
   \caption{Optical colours measured from combined flux-calibrated, dereddened spectra.  Solid boxes are ($B-V$), open boxes ($V-R$) and open triangles ($V-I$).}
\label{fig:gx_op_col}
\end{centering}
\end{figure}

All three colour indices increased from 1998 May to 1999 May (see Figure \ref{fig:gx_op_col}).  In this case the errors are smaller than the greatest difference between 1998 and 1999 suggesting that this trend is real.  There is, however, large changes between 1999 March 25 and 26.  Therefore we cannot determine exactly how much the colours change between 1998 and 1999.

\subsection{Emission Line Variations over Epochs}
\label{sec:elvoe}

Spectra from each epoch were combined and normalised to the continuum.  Low and high resolution data were separated and remain so for the rest of this analysis.  

The EW and FWHM of the strongest emission lines were measured by fitting single-Gaussian profiles using \textit{splot} in IRAF.  For each spectrum three or four measurements were made of each line varying the placement of the continuum.  The average EW and FWHM were calculated from these measurements and the errors determined from their scatter.  

\subsubsection{Equivalent Width}
\label{sec:gx_comb_epoch_ew}

\begin{figure}
\begin{centering}
\includegraphics[width=7cm,height=8cm]{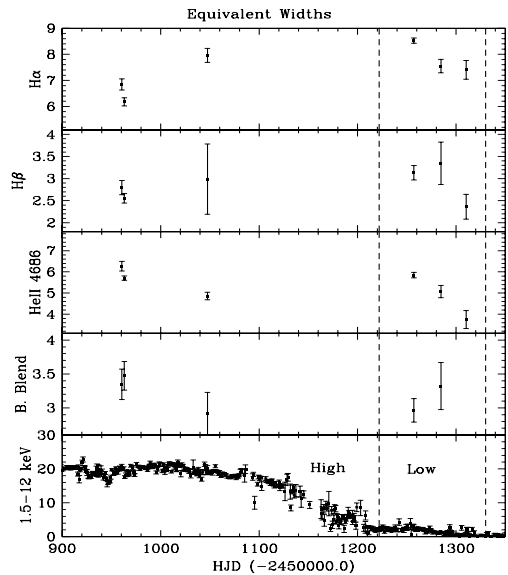}
   \caption{Equivalent widths (\AA) of the strongest emission lines in GX 339-4 from combined spectra of observed epochs.  Bottom panel shows soft X-rays for comparison. `High' and `Low' refer to the X-ray states.  }
\label{fig:eqw_xrays}
\end{centering}
\end{figure}

The EW of {\ha}, {\hb}, {\heii} and the {\bb} for each epoch are listed in Table \ref{tab:eqw_comb} and are plotted in Figure \ref{fig:eqw_xrays} together with the soft X-ray flux.  

\begin{table*}
\begin{center}
\caption{Equivalent widths (\AA) of the strongest emission lines in GX 339-4 from combined spectra of observed epochs. }
\label{tab:eqw_comb}
  \begin{tabular}{r|l|c|c|c|c}
\hline
Epoch & HJD & H$\alpha$ & H$\beta$ & HeII $\lambda$4686 & Bowen Blend \\
\hline
1998 May (low res.) & 2450960.61940 & 6.8 $\pm$ 0.2 & 2.8 $\pm$ 0.2 & 6.3 $\pm$ 0.2 & 3.3 $\pm$ 0.2 \\
1998 May (high res.) &  2450963.08762 & 6.2 $\pm$ 0.2 & 2.6 $\pm$ 0.1 & 5.7 $\pm$ 0.1 & 3.5 $\pm$ 0.2 \\
1998 August & 2451047.47701 & 8.0 $\pm$ 0.3 & 3.0 $\pm$ 0.8 & 4.9 $\pm$ 0.2 & 2.9 $\pm$ 0.3 \\
1999 March & 2451257.19183 & 8.5 $\pm$ 0.1 & 3.1 $\pm$ 0.2 & 5.9 $\pm$ 0.1 & 3.0 $\pm$ 0.2 \\
1999 April & 2451284.61028 & 7.6 $\pm$ 0.3 & 3.3 $\pm$ 0.5 & 5.1 $\pm$ 0.3 & 3.3 $\pm$ 0.3\\
1999 May & 2451310.24667 & 7.4 $\pm$ 0.4 & 2.4 $\pm$ 0.3 & 3.8 $\pm$ 0.4 & - \\
\hline
 \end{tabular}
\end{center}
\end{table*}

The 1998 May data show a decrease of EW in {\ha}, {\hb} and {\heii} from low resolution to high resolution data by $\sim$ 0.7, 0.2 and 0.6 \AA, respectively.  There is an increase in the {\bb} EW but the errors are quite large.  

The EW of {\ha} increased in 1998 August by $\sim$ 1.8 {\AA} while that of {\heii} and the {\bb} decreased by 0.8 and 0.6 \AA, respectively.

In the low state one clearly sees a decreasing trend of EW in {\ha}, {\hb} and {\heii} and an increase in the {\bb}, although the {\bb} again has large errors. 

We have seen that changes in the continuum slope are possible (see Figure \ref{fig:combdereddened}).  If the slope of the continuum becomes more negative (that is, bluer) then the continuum flux may decrease the equivalent width of {\heii} and increase that of {\ha}.  Hence the continuum may vary sufficiently to account for some of the changes seen here.  

The decrease in EW during the low state suggests that the line flux must be decreasing faster than the continuum.  

Without simultaneous photometry it cannot be said which of the continuum or line flux variations are the major causes to the EW changes over the year.  Analysis of future spectroscopic observations will benefit greatly if obtained simultaneously with photometry so that we may discern the cause of the EW changes.

\subsubsection{Full-Width Half Maximum}
\label{sec:gx_orbit_epoch_fwhm}

\begin{figure}
\begin{centering}
\includegraphics[width=7cm,height=8cm]{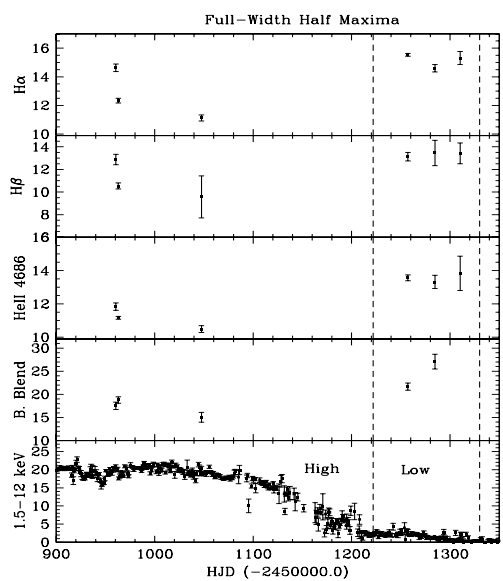}
   \caption{Full-width half maxima (\AA) of the strongest emission lines in GX 339-4 from combined spectra of observed epochs.  Bottom panel shows soft X-rays for comparison. }
\label{fig:fwhm_xrays}
\end{centering}
\end{figure}

The FWHM of the strongest emission lines are given in Table \ref{tab:fwhm_comb} and plotted in Figure \ref{fig:fwhm_xrays} with the soft X-ray flux.    

\begin{table*}
\begin{center}
\caption{Full-width half maxima ({\kms}) of the strongest emission lines in GX 339-4 from combined spectra of observed epochs. }
\label{tab:fwhm_comb}
  \begin{tabular}{r|l|c|c|c|c}
\hline
Epoch & HJD & H$\alpha$ & H$\beta$ & HeII $\lambda$4686 & Bowen Blend \\
\hline
1998 May (low res.) & 2450960.61940 & 667 $\pm$ 14 & 796 $\pm$ 31 & 755 $\pm$ 13 & 1130 $\pm$ 51 \\
1998 May (high res.) &  2450963.08762 & 564 $\pm$ 8 & 748 $\pm$ 19 & 717 $\pm$ 6 & 1210 $\pm$ 45 \\
1998 August & 2451047.47701 & 507 $\pm$ 9 & 592 $\pm$ 120 & 672 $\pm$ 13 & 976 $\pm$ 71 \\
1999 March & 2451257.19183 & 709 $\pm$ 5 & 808 $\pm$ 25 & 868 $\pm$ 12 & 1400 $\pm$ 52 \\
1999 April & 2451284.61028 & 667 $\pm$ 14 & 833 $\pm$ 68 & 851 $\pm$ 26 & 1750 $\pm$ 100 \\
1999 May & 2451310.24667 & 699 $\pm$ 23 & 826 $\pm$ 56 & 883 $\pm$ 64 & - \\
\hline
 \end{tabular}
\end{center}
\end{table*}

The FWHM decreases between the low and high resolution data of 1998 May in {\ha}, {\hb} and {\heii} by $\sim$ 100, 140 and 45 {\kms}, respectively.  In all lines the FWHM decreased from 1998 May to 1998 August then increased in 1999 by as much as $\sim$ 190 {\kms} for {\ha}.

The increase in FWHM from the high state to the low state is most noticeable in the line wings (Figure \ref{fig:comb_fwhm_hilob}).  This suggests that the line emitting regions are moving closer to the compact object from the high state to low state.  As stated above (\S \ref{sec:shls}), the disk may be cooling in the inner disk regions during the low state as there are fewer soft X-rays available to heat the disk.  Therefore regions closer to the compact object may emit Balmer lines during this time.  

It is worth noting that \citet{sor99} and \citet{wu01} saw no change in the line width of {\ha} between the high and low X-ray states but that of {\heii} increased with X-ray hardness.  In these observations a change in FWHM is detected in \textit{all} of the prominent emission lines, including {\ha}, from the high to low state.  It would seem, in this case at least, that all the strongest emission lines are correlated with X-ray hardness.  

The {\bb} FWHM is substantially larger than {\heii}.  This would imply that the {\bb} emission originates closer to the compact object.  The BFM, however, requires that {\heii} and the {\bb} emission originate from the same place since some of the HeII L$\alpha$ photons are absorbed and remitted by OIII and NIII.  Since it is a blend of several lines, the FWHM of the {\bb} may be 'artificially' increased. 

\subsection{{\hb} Absorption Feature}
\label{sec:hb_abs}

An interesting feature in the combined blue spectra is the redshifted absorption dip at {\hb} which is seen at all epochs.  No other emission lines exhibit such absorption.  Figure \ref{fig:gx-bluehilo} shows combined low-dispersion spectra.   As can be seen, {\ha} has a redshifted absorption during the low state.  When low and high dispersion spectra are combined (Figure \ref{fig:comb_fwhm_hilob}) the absorption dip is not evident.  Hence the absorption feature near {\ha} could be an artifact of combining low dispersion spectra only whereas the redshifted absorption component is present near {\hb} in all spectra.

\begin{figure}
\begin{centering}
\includegraphics[width=7cm,height=8cm]{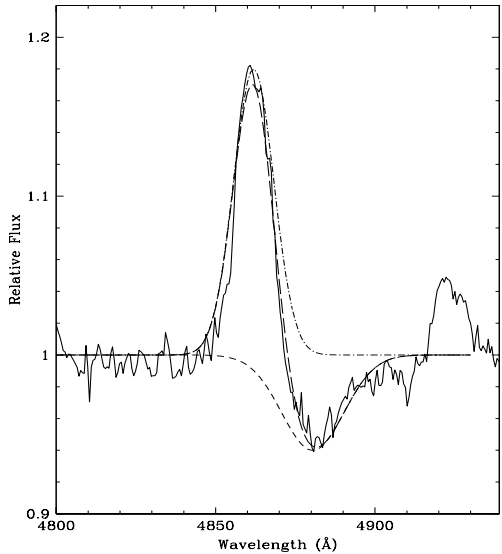}
   \caption{Redshifted absorption near {\hb}.  The solid line is the combined spectrum from all epochs centred on {\hb}.  The emission line to the right is HeI $\lambda$4922.  The dot-dash line is the Gaussian fit to {\hb} emission line.  The dashed line is the Gaussian fit to absorption trough.  The combined Gaussian is shown by the long-dashed line. }
\label{fig:hbeta_gauss}
\end{centering}
\end{figure}

Figure \ref{fig:hbeta_gauss} shows the combined spectra from all epochs expanded around {\hb}.  Several Gaussian fits were made to the {\hb} emission and absorption lines.  An example fit is also shown in Figure \ref{fig:hbeta_gauss}.  The Gaussian fits have the following properties:  {\hb} Gaussian: FWHM = 780 $\pm$ 200 {\kms}, EW = 2.67 $\pm$ 1.0 \AA, $\lambda_c$ = 4861.94 $\pm$ 0.03 \AA; absorption Gaussian:  FWHM = 1200 $\pm$ 280 {\kms}, EW = 0.90 $\pm$ 0.03 \AA, $\lambda_c$ = 4880.49 $\pm$ 0.14 \AA.  The errors shown were derived from the scatter in the measurements.

The same feature has been observed in V404 Cyg by \citet{cas93} who attribute it to a metallic blend of CrI, NiI and FeI at $\lambda$4889.  Casares et al. found the spectral type of the secondary in V404 Cyg to be G9$\pm$1 V to K0$\pm$4 III derived by comparison of CaI and FeI lines in the range $\lambda\lambda$ 6350-6530 to template star spectra.  No absorption lines were seen in this spectral range for GX 339-4.  Thus it seems unlikely that the observed absorption near {\hb} originates from the secondary star.   

If the absorption line is an inverted P-Cygni profile then we would expect to also see it in {\ha}.  As {\ha} does not convincingly show a redshifted absorption feature in all spectra this explanation is also doubtful.  

Future observations should endeavour to see if this feature is still present near {\hb} and whether {\ha} exhibits the same or otherwise.  Obtaining higher SNR ratio data may enable us to ascertain the source of this absorption.

\section{Analysis of Individual Spectra}
\label{sec:indiv_spec}

\begin{figure}
\begin{centering}
\includegraphics[width=7cm,height=8cm]{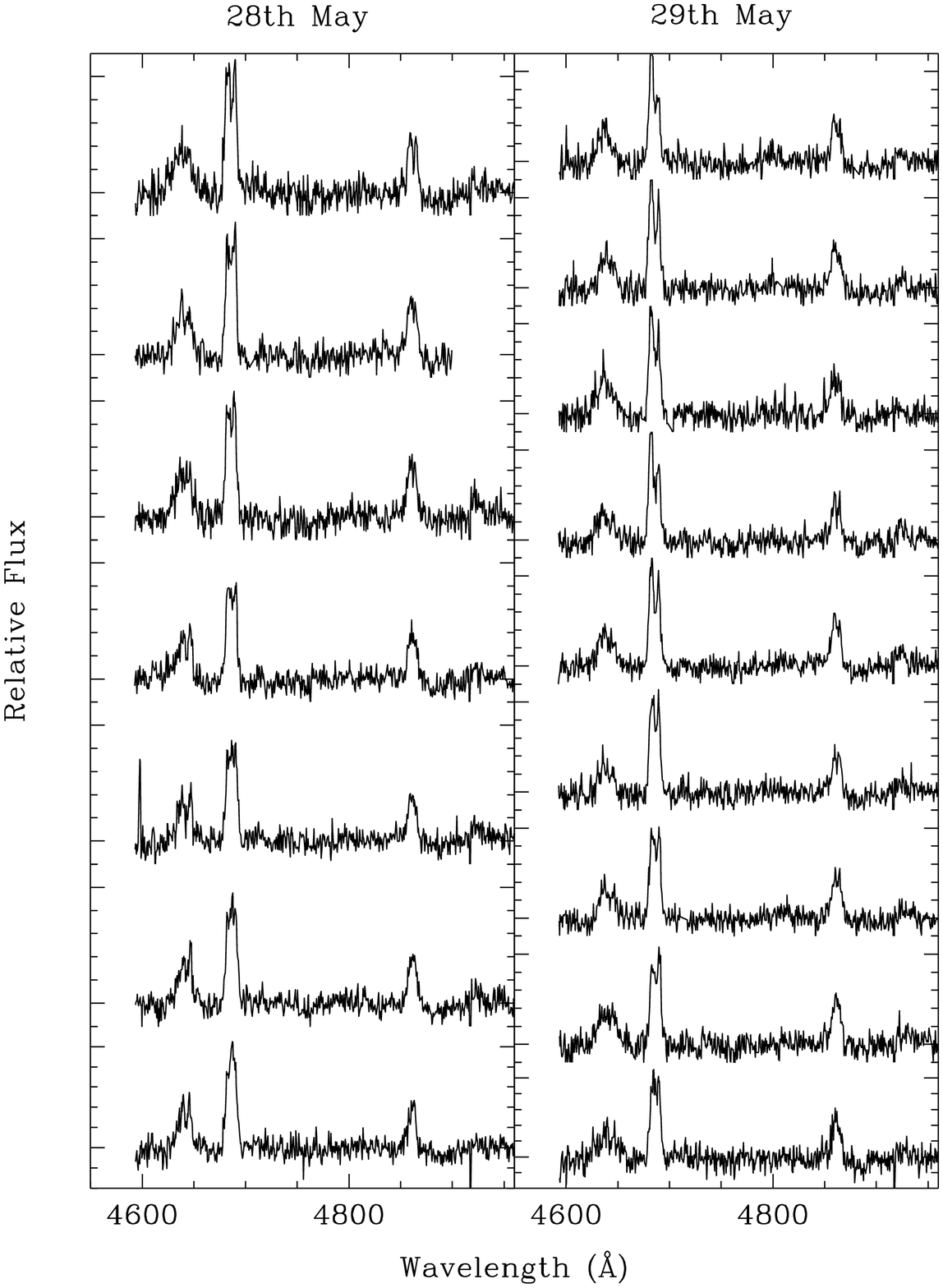}
\includegraphics[width=7cm,height=8cm]{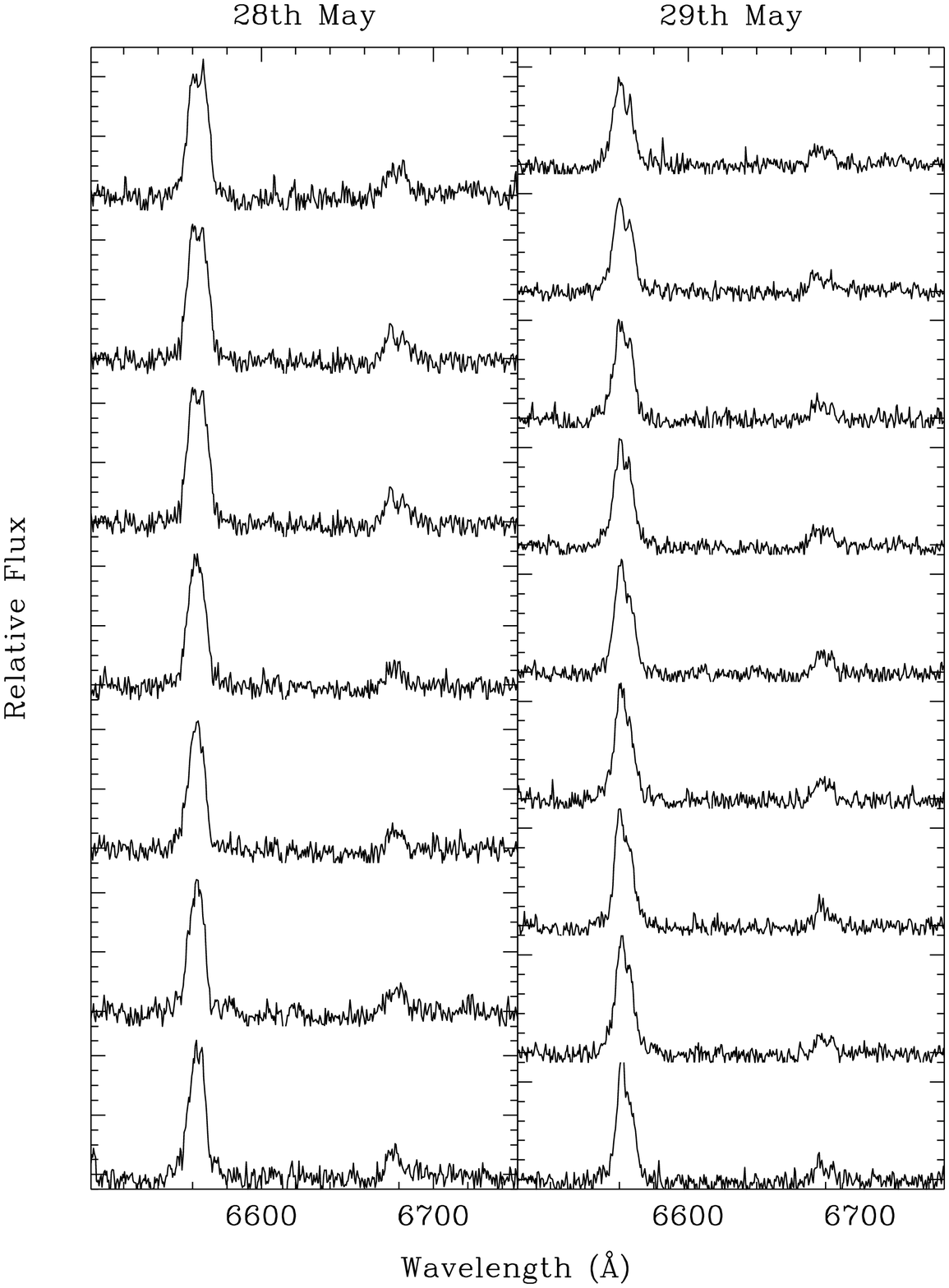}
   \caption{Individual high resolution blue (\textit{top}) and red (\textit{bottom}) spectra of GX 339-4 on 1998 May 28 and 29.  The emission lines visible from left to right are, in the blue spectra, {\bb}, {\heii} and {\hb} and, in the red spectra, {\ha} and HeI $\lambda$6678.}
\label{fig:ind2_may98b}
\end{centering}
\end{figure}

\begin{figure}
\begin{centering}
\includegraphics[width=7cm,height=8cm]{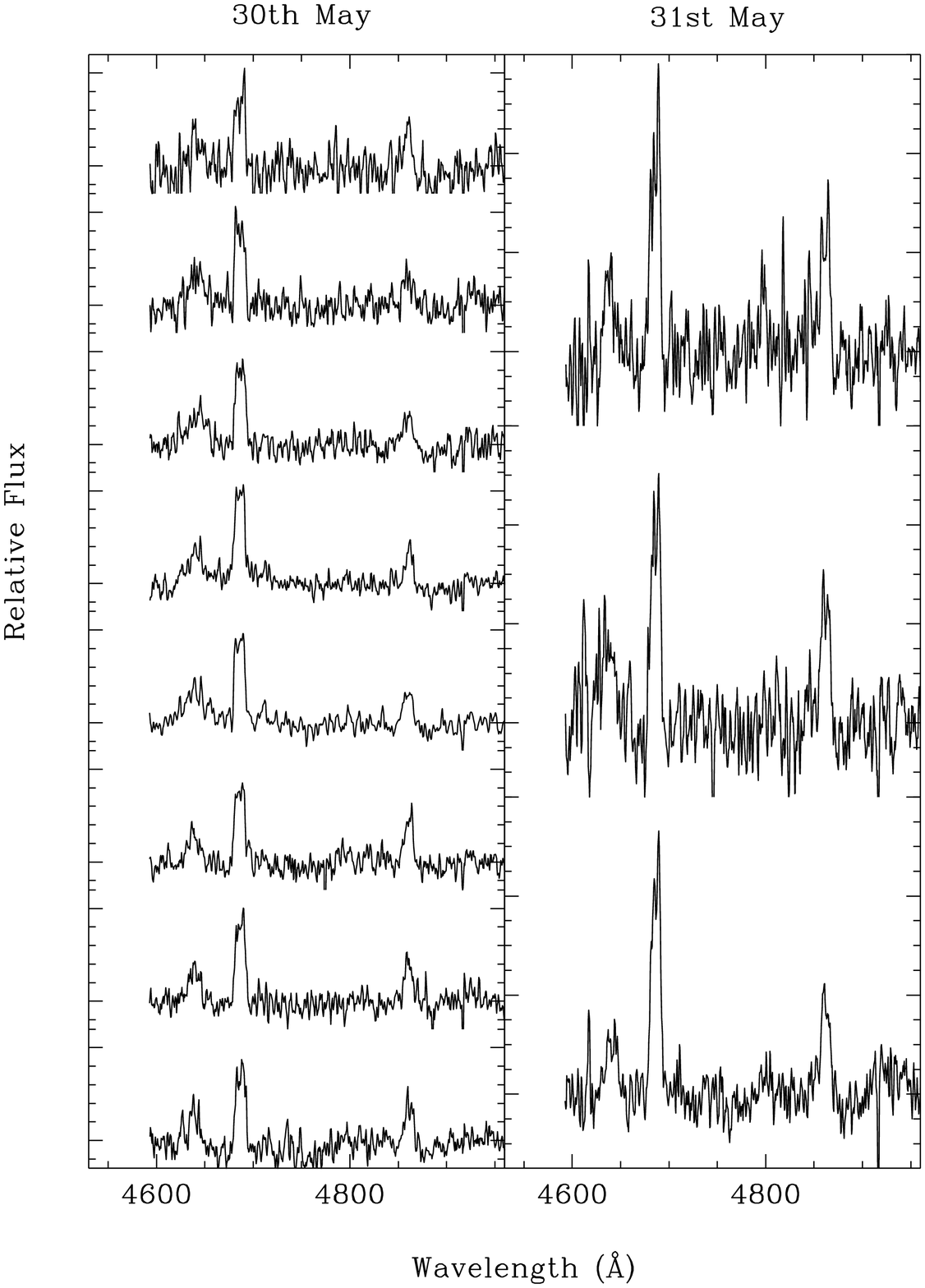}
\includegraphics[width=7cm,height=8cm]{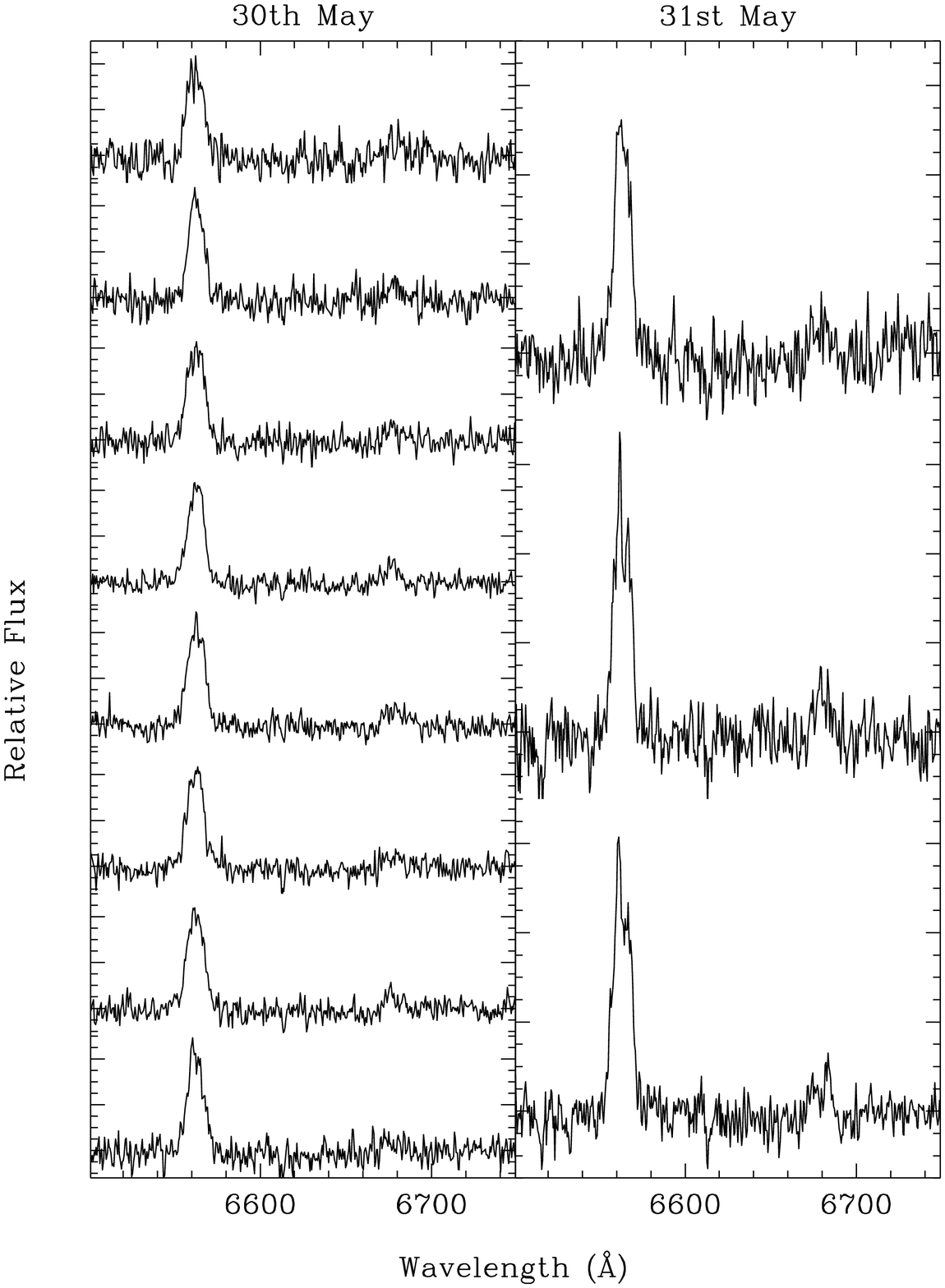}
   \caption{As for Figure \ref{fig:ind2_may98b} for 1998 May 30 and 31.}
\label{fig:ind3_may98b}
\end{centering}
\end{figure}

An example of low and high resolution individual spectra are plotted in Figures \ref{fig:ind2_may98b} and \ref{fig:ind3_may98b}.  Inspection of the spectra by eye show that the line profiles vary on a timescale of hours.  One can see in particular the peak flux in {\heii} vary on the blue side on May 29.  The {\heii} profile on May 28 looks more symmetrical and the double-peaks are not as clear.  In {\ha} it is difficult to discern two well-discerned peaks at any time although one may argue that they are present in the first few spectra on May 29 and perhaps May 31.  {\ha} is more round-topped and symmetric on May 28 then becomes asymmetric on the blue side on May 29.  The spectra repeat this profile behaviour the following two nights (Figure \ref{fig:ind3_may98b}).  Since the profile characteristics repeat themselves this strongly suggests that the profile variations are related to the orbital period. 

The following analysis attempted to extract quantitative information on the changes in the line profiles to see if they indeed corresponded to an orbital period and whether this agreed with the published orbital period of CCHT92.

At this point we note that the profile shape changes from 1998 to 1999 when we begin to see red-sided asymmetry.  This is studied in more detail in Paper II.  Plots of individual spectra from other epochs can be found in \citet{bux02}.

The analysis was conducted in three ways.  

First a single Gaussian profile was fitted to the most prominent emission lines using $\textit{splot}$ in IRAF.  Several measurements were taken of each line in each spectrum varying the placement of the continuum.  The average of the EW, FWHM and line centres (leading to radial velocities) were calculated from these measurements.  Single-Gaussian profiles do not perfectly fit the line profiles, particularly in the central regions of the emission line.  As the profiles fitted the line wings well we were satisfied that the FWHM and line centres were measured accurately.  To test how well the EW was measured, we used the ``e'' option in \textit{splot} within IRAF and found that there were virtually no differences between the two methods.     

The second method of analysis used double-Gaussians to fit profiles in cases where the two peaks were well discerned.  Several measurements were made varying the continuum and Gaussian centre placements.  The average EW, FWHM and radial velocities of both Gaussian components were calculated.  

Another way of quantifying the line profile variations is by taking the ratio of fluxes from the blue (or violet) and red sides of the emission line, also known as the V/R ratio.  The advantage of this method is that no assumptions are made on the shape of the line profile.  The V/R ratio was calculated for H$\alpha$ and {\heii}.  The dividing line between the blue and red fluxes was taken to be the rest wavelength, that is 6562.80 {\AA} for {\ha} and 4685.68 {\AA} for {\heii}.  

A period search was conducted on the EW, FWHM, radial velocities and V/R ratios.  The search was performed over a restricted range of frequencies to eliminate possible aliases.  The frequencies searched over ranged from three times the exposure of the individual spectra to one third of the epoch coverage.  

Due to the size of the dataset it is not possible to show every single result of the analysis.  Hence, we will first discuss briefly the datasets for which the results are not presented here.  

In only a few cases, epochs did not have sufficient temporal coverage such that the published orbital period was not sampled in the periodogram.  This was always true for the low resolution 1998 May dataset.  

The SNR of individual spectra at many epochs was quite low.  The SNR of high resolution spectra in 1998 May was 50-60 compared to 10-20 for high resolution spectra in 1998 August.  The errors in EW, FWHM and radial velocities were too large to give any meaningful results in the low SNR cases. 

Analysis of the double-gaussian fits did not show any modulation in the radial velocities, EW, FWHM or V/R ratios.
 
Most epochs which had a SNR of $\sim$ 40 did not have a sufficient number of spectra to cover all orbital phases.

Hence, \textit{the discussion of results in the following sections concentrate on those from the high resolution data of 1998 May 28-31}.  This dataset provided the best temporal coverage, had the highest SNR and had the most number of spectra.  

For results on analysis of other epochs the reader is referred to \citet{bux02}.

\subsection{Radial Velocities}
\label{sec:indiv_rv}

\begin{figure}
\begin{centering}
\includegraphics[width=7cm,height=8cm]{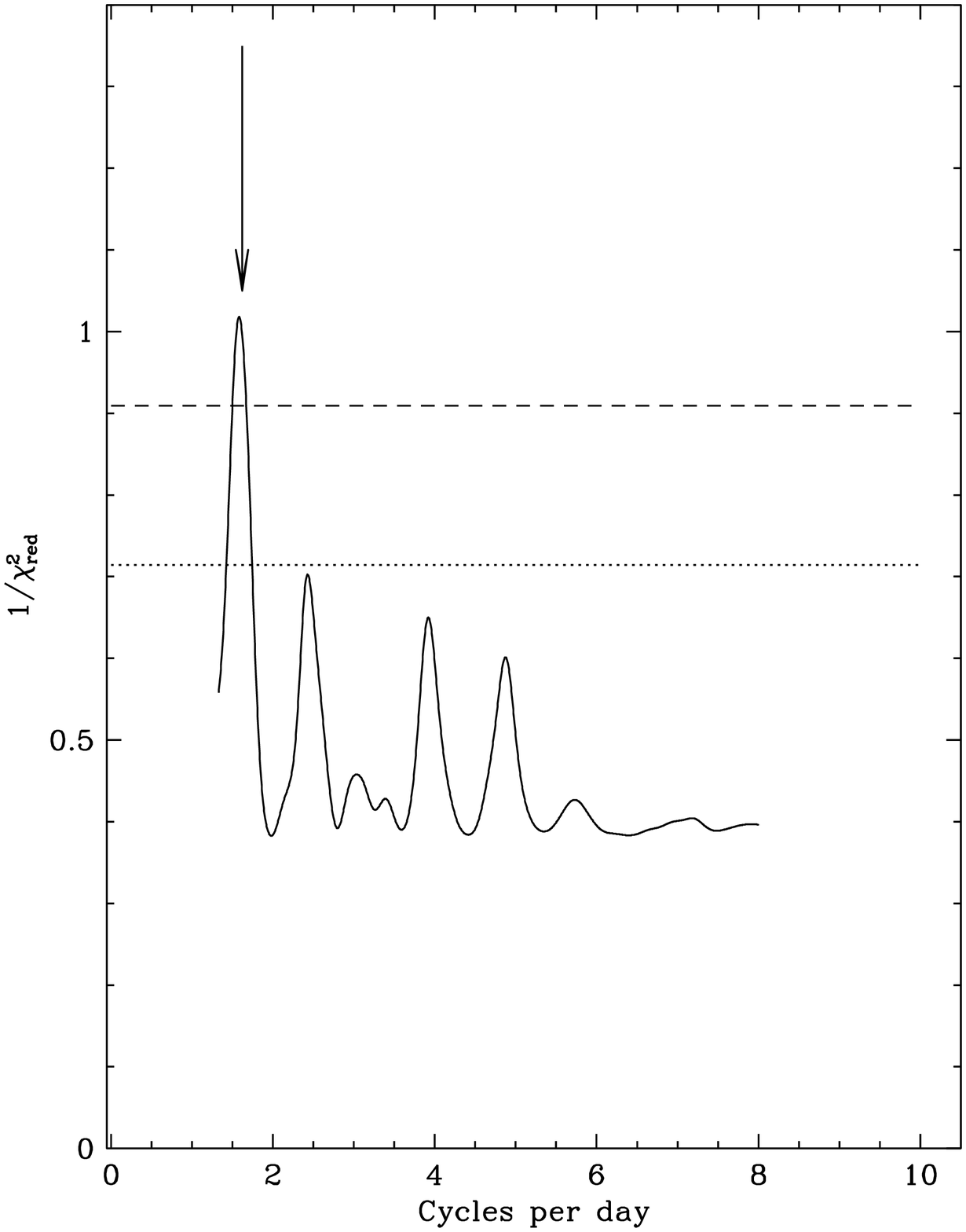}
\includegraphics[width=7cm,height=8cm]{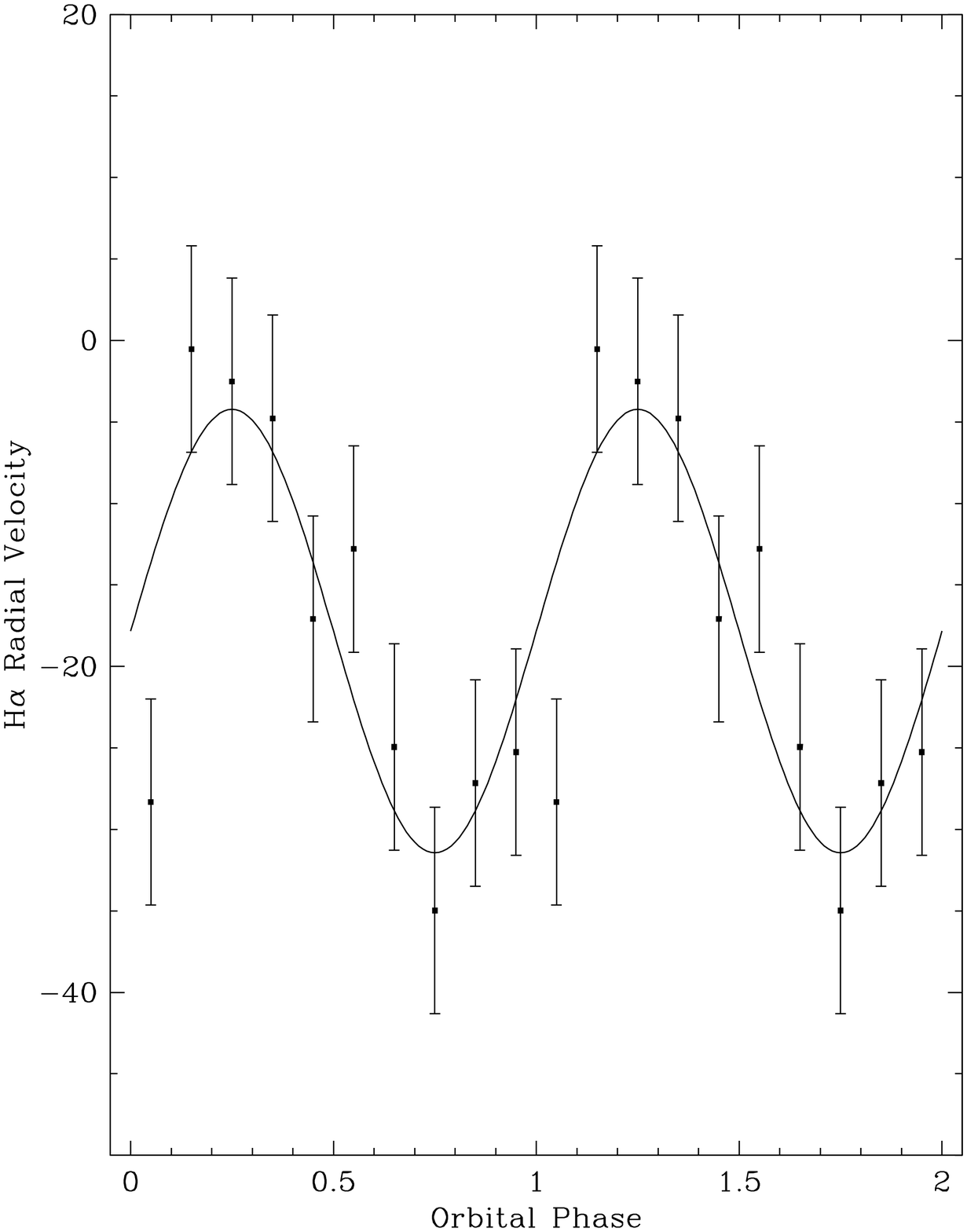}
    \caption{$Top$:  Periodogram of {\ha} radial velocities measured from single-Gaussian line profile fits.  Data is high resolution spectra from 1998 May 28-31.  Arrow shows where the published period of 14.86 hours lies.  $y$-axis is the normalised, reduced $\chi^2$.  Long dashed line shows 99\% confidence level level and dotted line the 68\% confidence level.  $Bottom$:  Radial velocities of H$\alpha$ measured from single-Gaussian profile fits phased on the 14.86 hours period.  Points were grouped into 0.10 phase bins.  Solid curve is best-fit sinusoid derived from the least-squares sine fit to the binned data.  Errorbars are the standard deviation derived from the least-squares fit.}
\label{fig:per_svel_halpha}
\end{centering}
\end{figure}

\begin{figure}
\begin{centering}
\includegraphics[width=7cm,height=8cm]{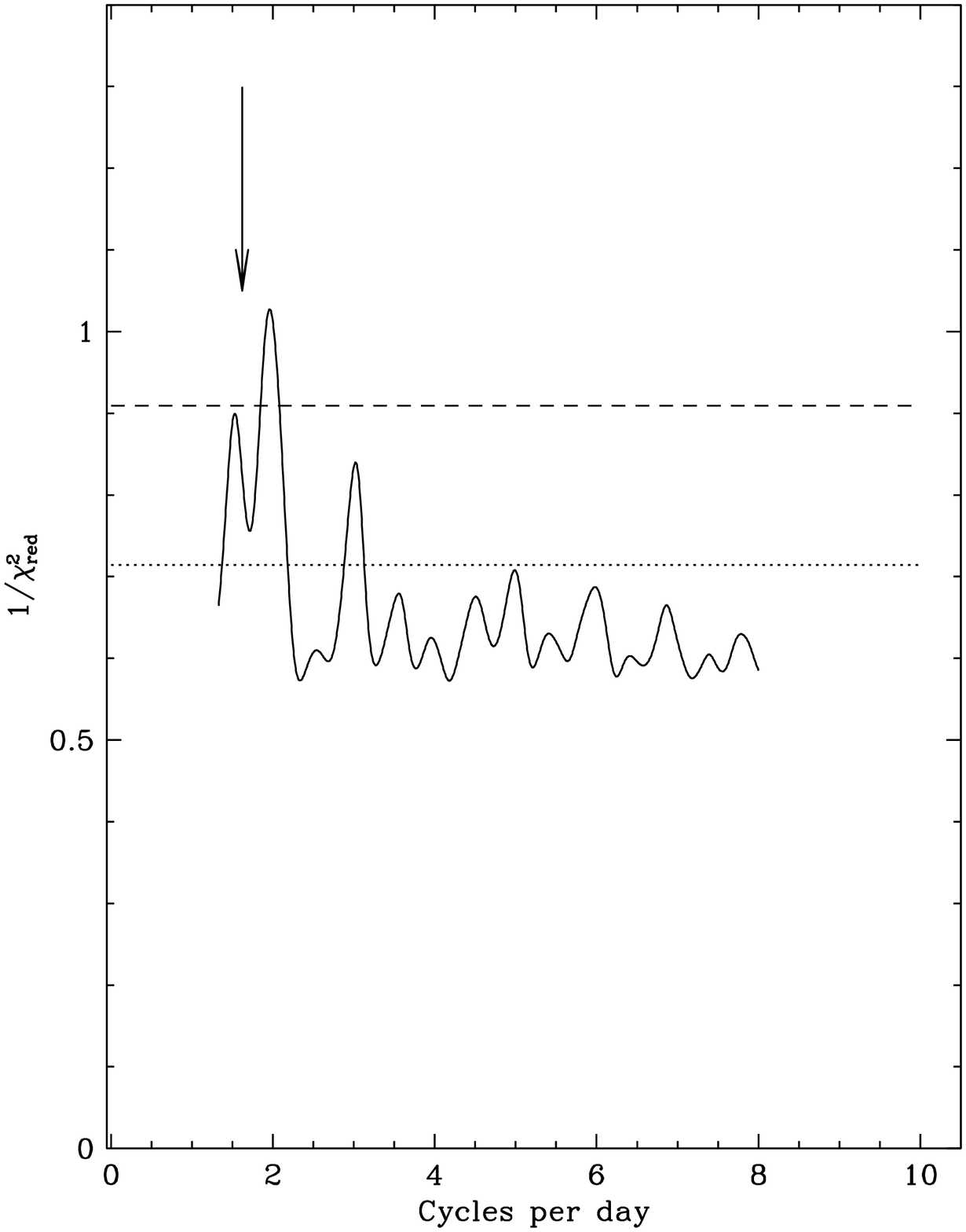}
\includegraphics[width=7cm,height=8cm]{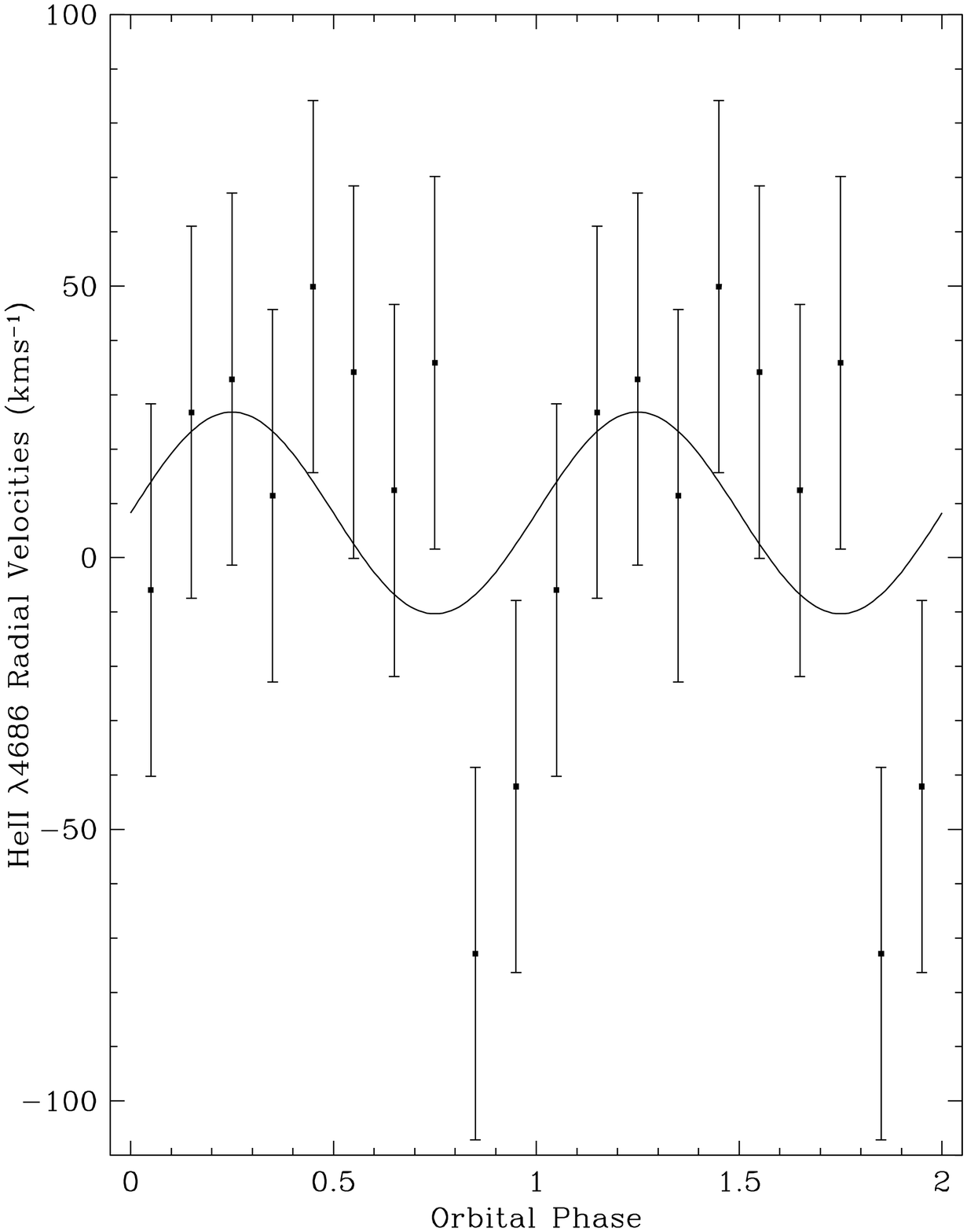}
    \caption{As for Figure \ref{fig:per_svel_halpha} for {\heii}. }
\label{fig:per_svel_heii4686}
\end{centering}
\end{figure}

\begin{figure}
\begin{centering}
\includegraphics[width=7cm,height=8cm]{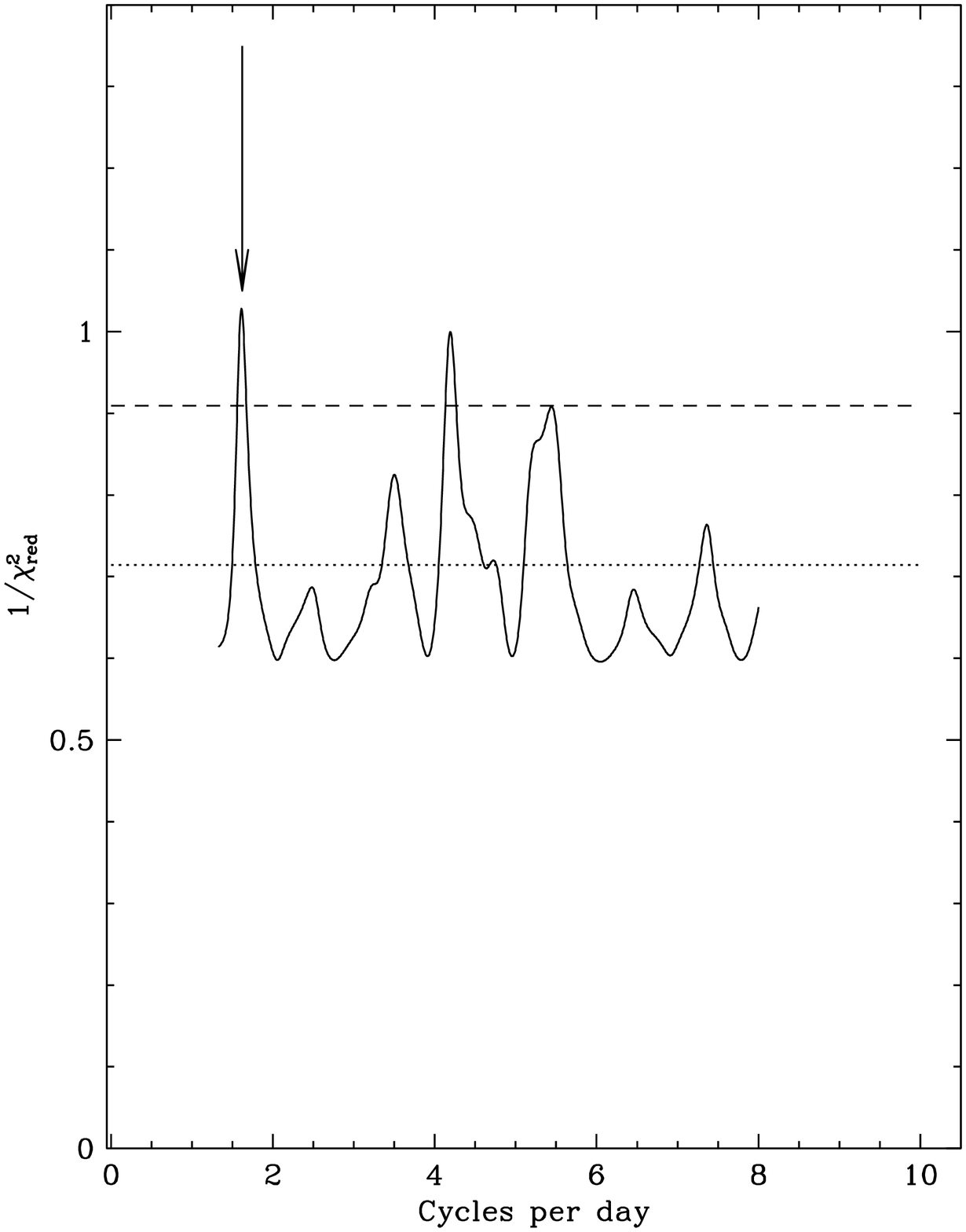}
\includegraphics[width=7cm,height=8cm]{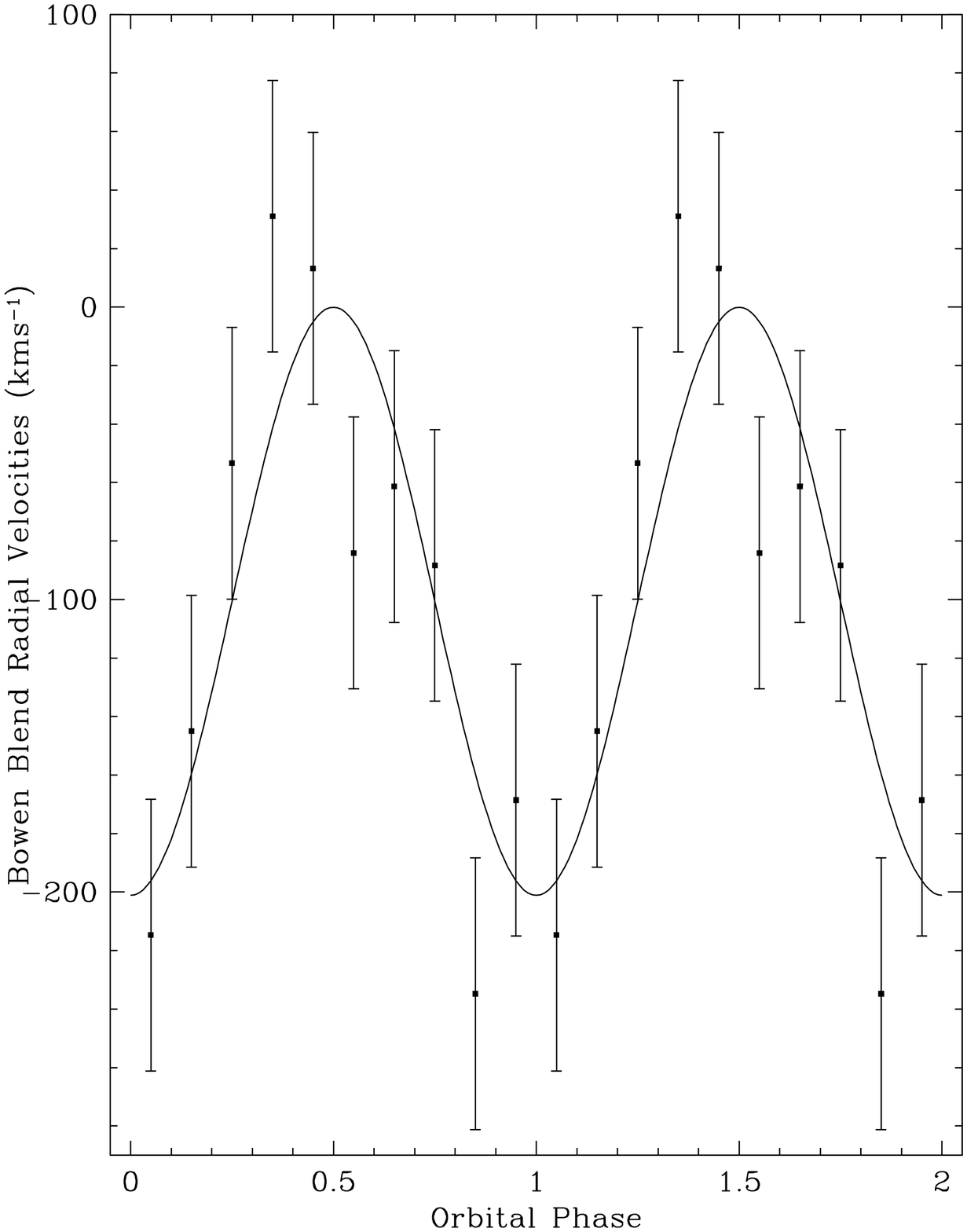}
    \caption{As for Figure \ref{fig:per_svel_halpha} for the {\bb}. }
\label{fig:per_svel_bblend}
\end{centering}
\end{figure}

Periodograms created from the {\ha}, {\heii} and the {\bb} radial velocities measured via single-Gaussian profiles are shown in Figures \ref{fig:per_svel_halpha} to \ref{fig:per_svel_bblend}.  The 68\% and 99\% confidence levels shown were calculated using the method outlined in \citet*{lam76} for the two-parameter model case.  

There is a strong peak in all periodograms at the 99\% confidence level.  The peak for {\ha} falls at the published orbital period.  The periodogram of {\heii}, however, has a broad double-peak around the published orbital period.  The {\bb} shows a strong peak at the published orbital period.  There are, however, peaks at other periods which are as significant.

Before folding the radial velocities on the orbital period it was first necessary to calculate a new ephemeris.  The ephemeris given by CCHT92 is P$_{orb}$ = 0.61916$\pm$0.0027 days and $T_{o}$ (JD, minimum light) = 2447000.539 $\pm$ 0.015.  The error on the period is large enough to render the ephemeris useless after $\sim$ 140 days and, therefore, could not be used to correctly phase this dataset.  

A new ephemeris was determined from the radial velocities of H$\alpha$ which were measured from single-peaked Gaussians.  To obtain the phase shift a period search was conducted on the data with the period restricted to the published period of 0.61916 days.  The amount of time corresponding to this phase shift was subtracted from the time of the first observation, giving a new $T_{o}$ (JD) = 2450961.8773 $\pm$ 0.0078.  The error in $T_{o}$ was calculated by conducting the period search again on the period with the error subtracted and added and finding $T_{o}$ in both cases.  The error given is the larger of the two.  

The orbital period could not be refined as the errors on the radial velocities over the various epochs were too large.

Phase 0.0 corresponds to the time when the radial velocities cross from the blue to the red.  Since the majority of the emission line flux originates from the accretion disk as indicated by the double-peaked profiles \citep{hua72, sma69} the radial velocity variations most likely relate to the disk.  Therefore, phase 0.0 corresponds to the inferior conjunction of the primary.  One must use caution, however, as line asymmetries caused by, say, a hotspot may shift the line centroid and give a false indication of the time of maximum and minimum of the radial velocities.  This would require more detailed analysis involving line profile modelling which is pursued in Paper II.  For now we take phase 0.0 to be the point at which the accretion disk is between the observer and the secondary star. 

The radial velocities of {\ha}, {\heii} and the {\bb} were folded on the published orbital period using the new $T_o$.  The results are also shown in Figures \ref{fig:per_svel_halpha} to \ref{fig:per_svel_bblend}.  Points were binned by 0.10 in phase.  The binned data were subjected to a least-squares sine fit.  The solid curves in the figures show the best-fit sinusoid.  Their parameters are listed in Table \ref{tab:rv_bestfit}.  The errors on the semi-amplitudes and systemic velocities were obtained from the fit.  The errorbars on the radial velocities are the standard deviations obtained from the best-fit sinusoids.    

\begin{table}
\begin{center}
\caption{Parameters of best-fit sinusoid curves to radial velocities of emission lines. }
\label{tab:rv_bestfit}
  \begin{tabular}{l|c|c}
\hline
Line & $K_1$ & $\gamma$ \\
 & ({\kms}) & ({\kms}) \\
\hline
{\ha} & 14 $\pm$ 3 & -18 $\pm$ 2 \\
{\heii} & 20 $\pm$ 20 & 8 $\pm$ 10 \\
{\bb} & 30 $\pm$ 20 & -10 $\pm$ 20 \\
\hline
 \end{tabular}
\end{center}
\end{table}

The {\ha} radial velocity curve shows clear modulation on the published orbital period of 14.86 hours.  The maximum is at phase 0.25 and the minimum at phase 0.75.  

The radial velocity curve of {\heii} is not as well defined as {\ha} and is obviously why the periodogram did not show an isolated peak at the published orbital period.  Consequently the best-fit sine curve does not describe the data well.  

The {\bb} shows a much cleaner radial velocity curve than {\heii}.  The maximum, however, is at phase 0.5 and minimum at phase 0.0.  This is 0.25 out of phase with respect to the {\ha} radial velocities.  In addition, the velocity amplitude is much larger than {\ha} and {\heii}.  Since the {\bb} is a blend of several lines which may vary in strength with time it is not a reliable velocity tracer.  

Previous studies which use emission lines to determine $\gamma$ show that it is sensitive to distortions in the line profile and cannot be relied upon \citep{oro94, gar96}.  Since the lines are obviously asymmetric the $\gamma$ derived here will not be discussed further but have been shown in Table \ref{tab:rv_bestfit} for completeness.

CCHT92 derived $K_1$ = 78 $\pm$ 13 {\kms} from the {\ha} radial velocity data obtained by \citet{cow87}.  This value is much higher than that obtained here.  A lower radial velocity amplitude decreases the mass function and increases the upper limit to the compact object mass.  Using $K_1$ = 14 {\kms} from {\ha}, and $P_{orb}$ = 14.86 hours, the mass function is $f(M)$ = 2 x $10^{-4}$ {\msun}.  Assuming $M_2$ = 0.7 {\msun} (K5 V star) and $i \le 70^o$, $M_1 \le 40$ {\msun}.  If $i = 15^o$ \citep{wu01} then $M_1 = 5$ {\msun}.  This would imply that the compact object in GX 339-4 is a black hole.  One must use caution, however, as previous studies have shown that using emission lines to measure $K_{1}$ may not be accurate \citep{wad85}.  In addition, the binary inclination is not well constrained although it is likely to be low.

\subsection{Equivalent Width}
\label{sec:gx_indiv_ew}

\begin{figure}
\begin{centering}
\includegraphics[width=7cm,height=8cm]{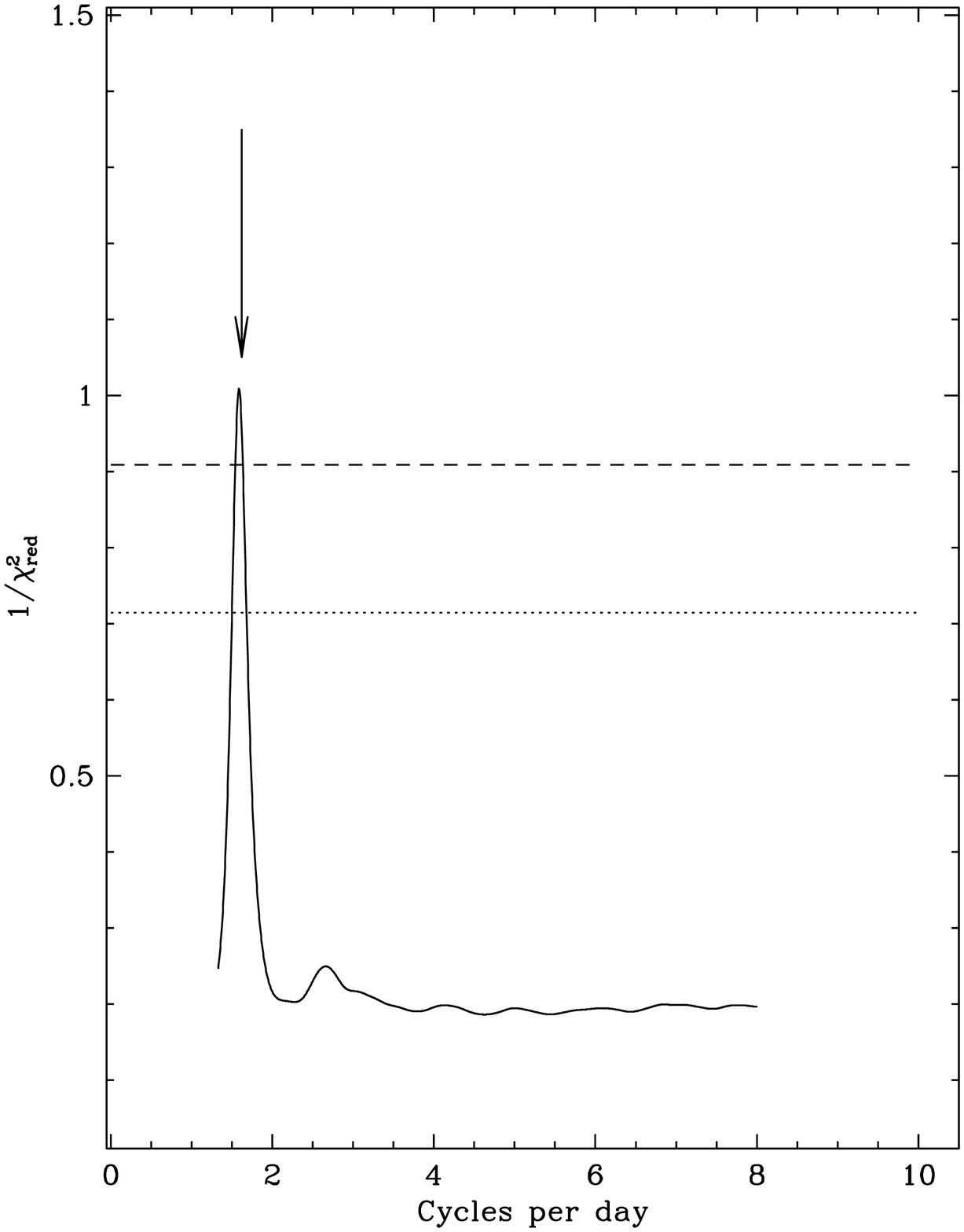}
\includegraphics[width=7cm,height=8cm]{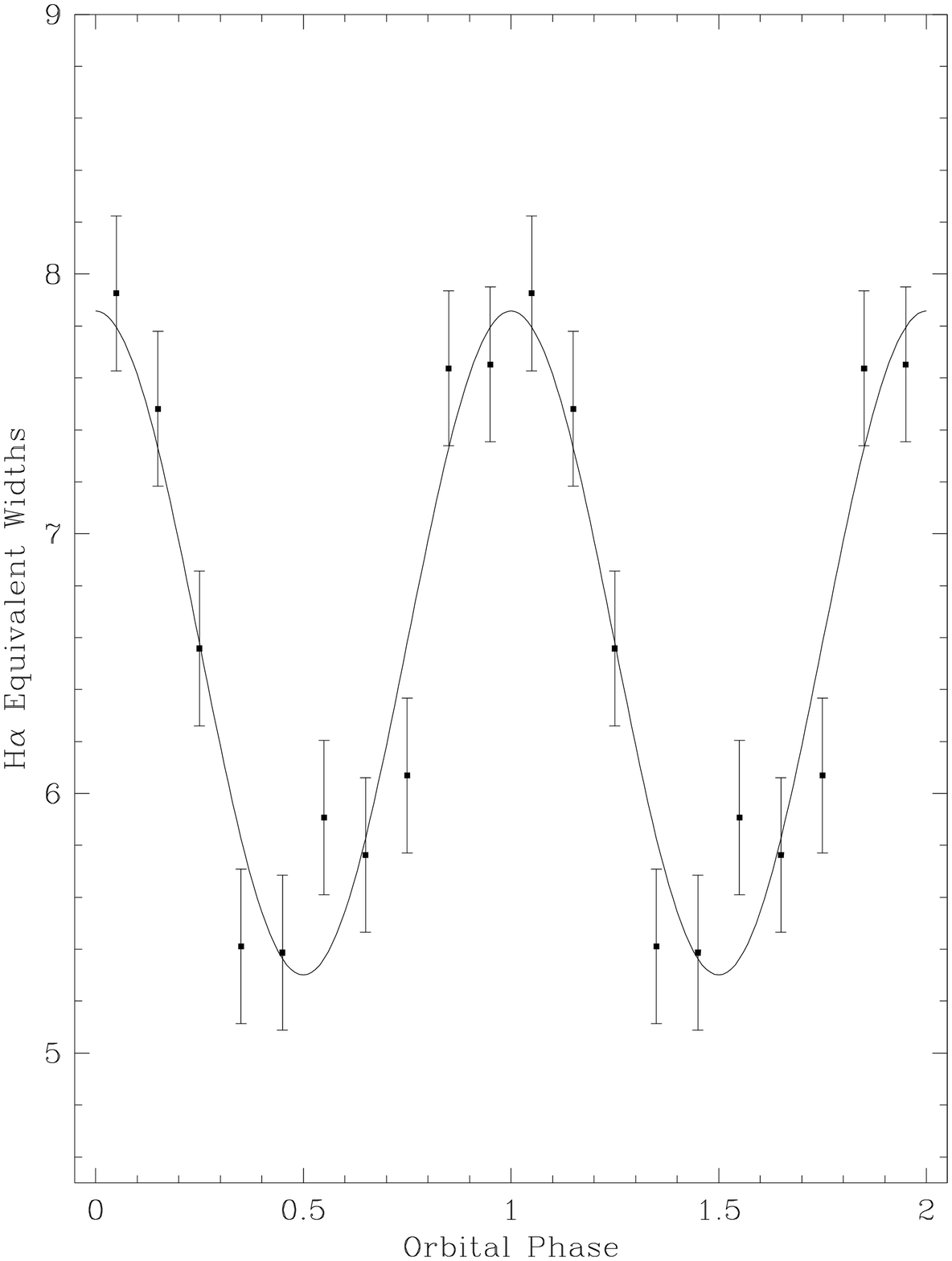}
    \caption{$Top$: Periodogram of {\ha} EW measured from single-Gaussian profile fits.  Data is high resolution spectra from 1998 May 28-31.  Arrow shows where the published period of 14.86 hours lies.  $y$-axis is the normalised, reduced $\chi^2$.  Long dashed line shows 99\% confidence level level and dotted line the 68\% confidence level.  $Bottom$:  EW of {\ha} folded on the published period.  Data points were binned by 0.10 in phase.  Solid curve is the best fit found from the least-squares sine fit to binned data.  Errorbars are the standard deviation derived from the least-squares fit.}
\label{fig:halpha_eqw_hmay98}
\end{centering}
\end{figure}

\begin{figure}
\begin{centering}
\includegraphics[width=7cm,height=8cm]{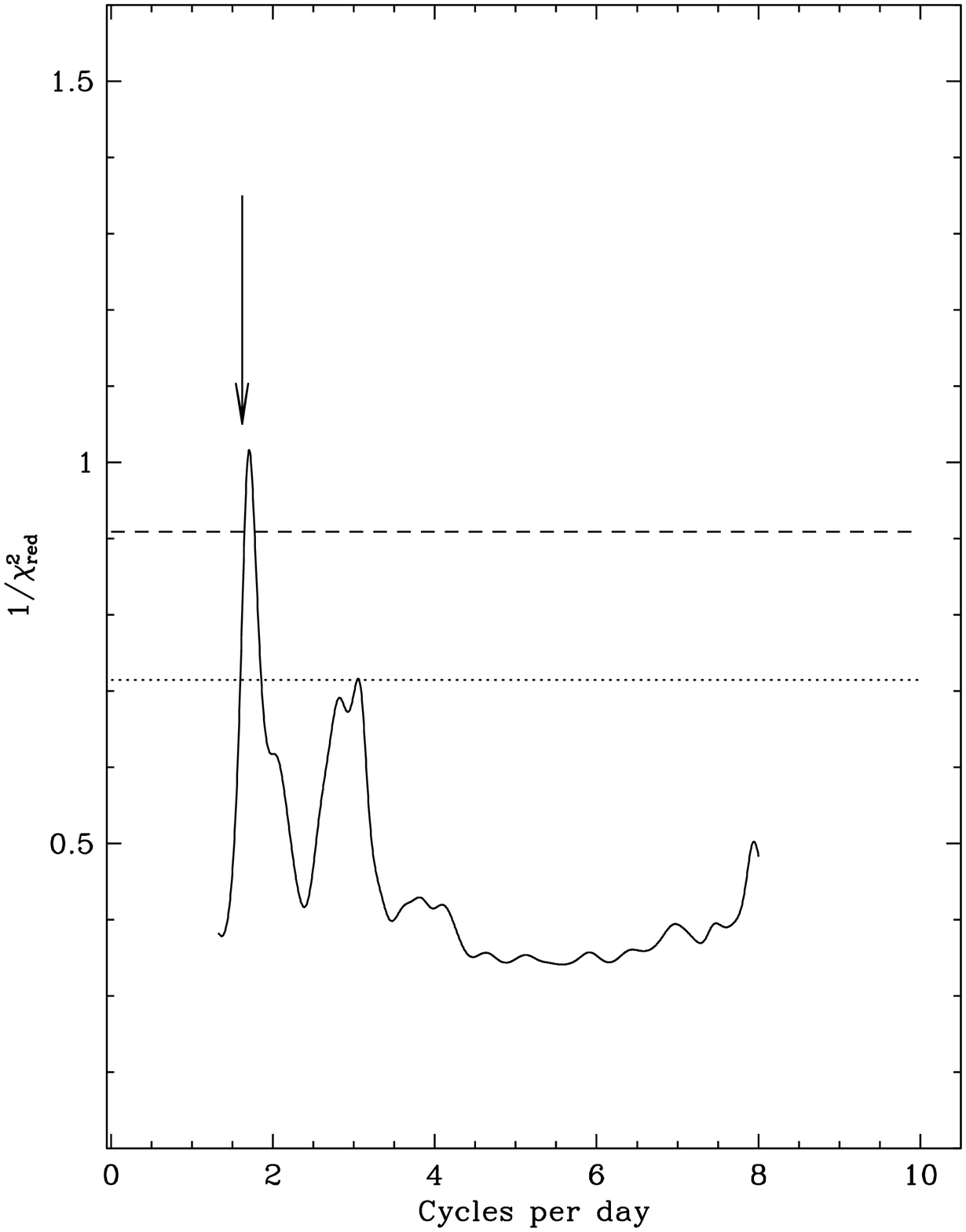}
\includegraphics[width=7cm,height=8cm]{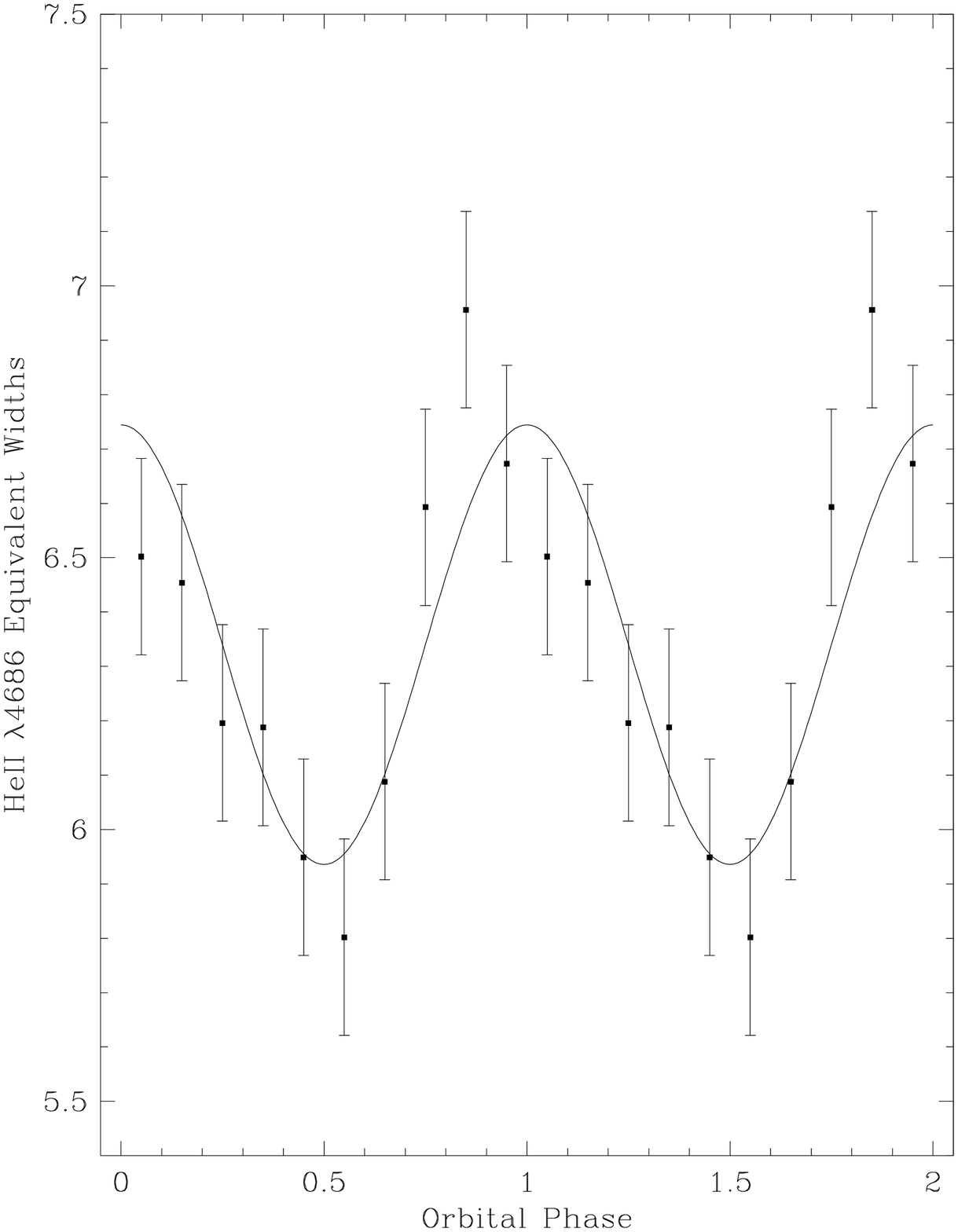}
    \caption{As for Figure \ref{fig:halpha_eqw_hmay98} for {\heii}.}
\label{fig:heii_eqw_hmay98}
\end{centering}
\end{figure}

Periodograms of the EW for {\ha} and {\heii} are shown in Figures \ref{fig:halpha_eqw_hmay98} and \ref{fig:heii_eqw_hmay98}.  Both cases show peaks at the 99\% confidence level at the published orbital period.  The {\bb} periodogram did not show any significant peaks \citep[see][]{bux02}. 

The EW of {\ha} and {\heii} were folded on the orbital period using $T_o$ calculated in \S \ref{sec:indiv_rv}.  Points were binned by 0.10 in phase.  Figures \ref{fig:halpha_eqw_hmay98} and \ref{fig:heii_eqw_hmay98} convincingly show a modulation of the EWs on the orbital period for both lines.  The sinusoid curves shown in the figures are the best fit to the data found from a least-squares sine fit.  Their parameters and errors are given in Table \ref{tab:eqw_bestfit}.  The errors on the EW are the standard deviations calculated from the least-squares sine fit.

\begin{table}
\begin{center}
\caption{Parameters of best-fit sinusoid curves to the EW of {\ha} and {\heii}. }
\label{tab:eqw_bestfit}
  \begin{tabular}{l|c|c}
\hline
Line & Semi-amplitude & Mean \\
 & (\AA) & (\AA) \\
\hline
{\ha} &  1.3 $\pm$ 0.2 & 6.6 $\pm$ 0.1 \\
{\heii} & 0.4 $\pm$ 0.1 & 6.2 $\pm$ 0.1 \\
\hline
 \end{tabular}
\end{center} 
\end{table}

In both cases the EW modulation has a maximum at phase 0.0 and minimum at phase 0.5.  This is 0.25 out of phase with respect to the {\ha} radial velocity curve.  In other words, the {\ha} and {\heii} EWs peak when the accretion disk is between the observer and the secondary star.  

\citet{gar96} observed an {\ha} EW modulation in GRO J0422+32 with a broad minimum at phase 0.5.  It peaked when {\ha} emission line radial velocities crossed from the blue to the red side and when the secondary star velocities crossed from red to blue.  Garcia et al. attributed the EW changes to an asymmetrically emitting disk which may also be asymmetric geometrically.  The observations were obtained during X-ray outburst and decline.

From comparison to Garcia et al.'s study of GRO J0422+32 we see the same behaviour in GX 339-4.  Hence, the EW variations in GX 339-4 may also be due to an asymmetric accretion disk.  Investigation of the line profile shape over the orbit may help in determining the source of the EW variations.  This is pursued in Paper II.

\subsection{Full-Width Half Maximum}

The periodograms of the FWHM measured from single-Gaussian fits to {\ha}, {\heii}, {\hb} and the {\bb} showed no significant peaks \citep[see][]{bux02}.  Hence, the FWHM were not considered any further.

\subsection{V/R Ratios}
\label{sec:vr_analysis}

\begin{figure}
\begin{centering}
\includegraphics[width=7cm,height=8cm]{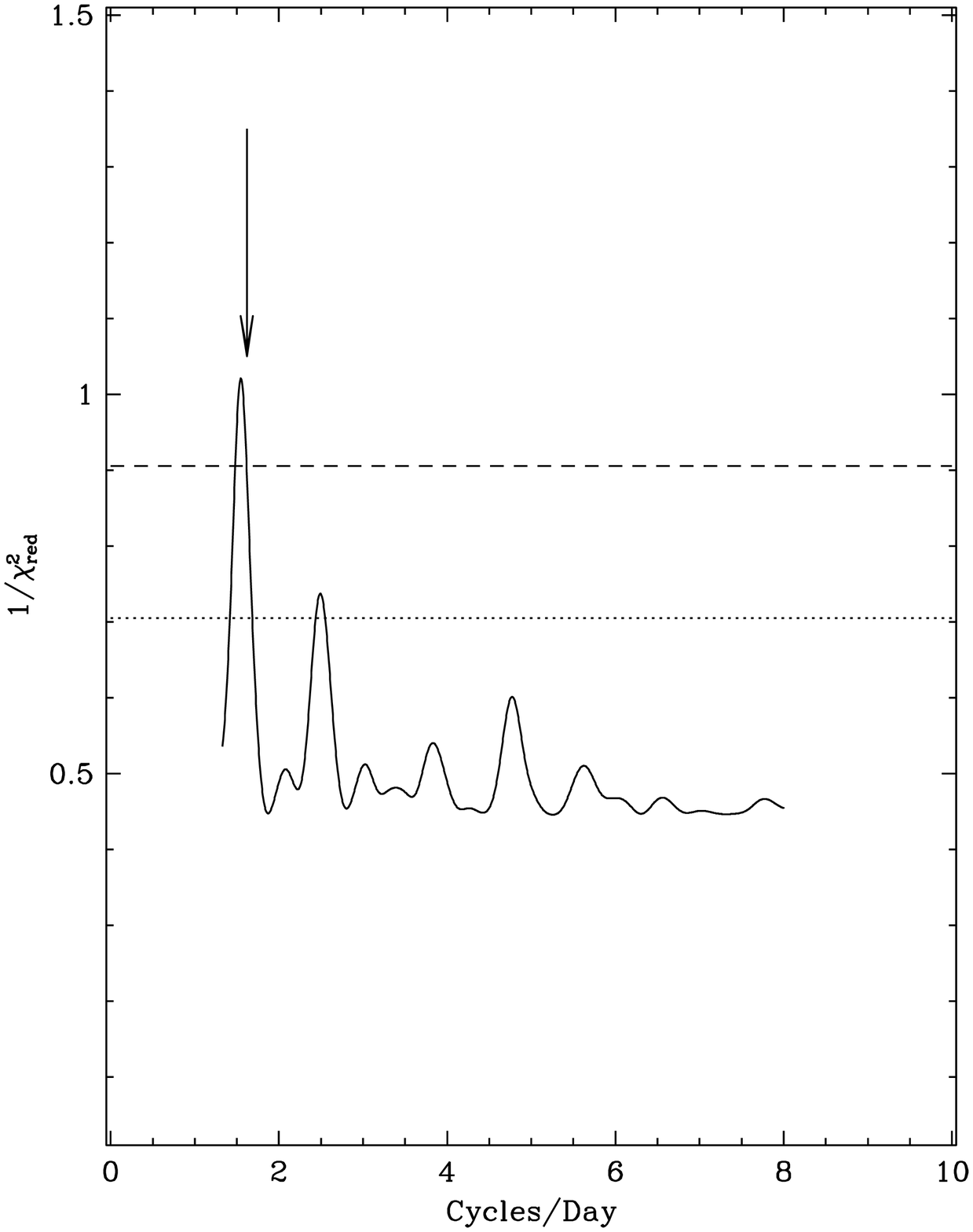}
\includegraphics[width=7cm,height=8cm]{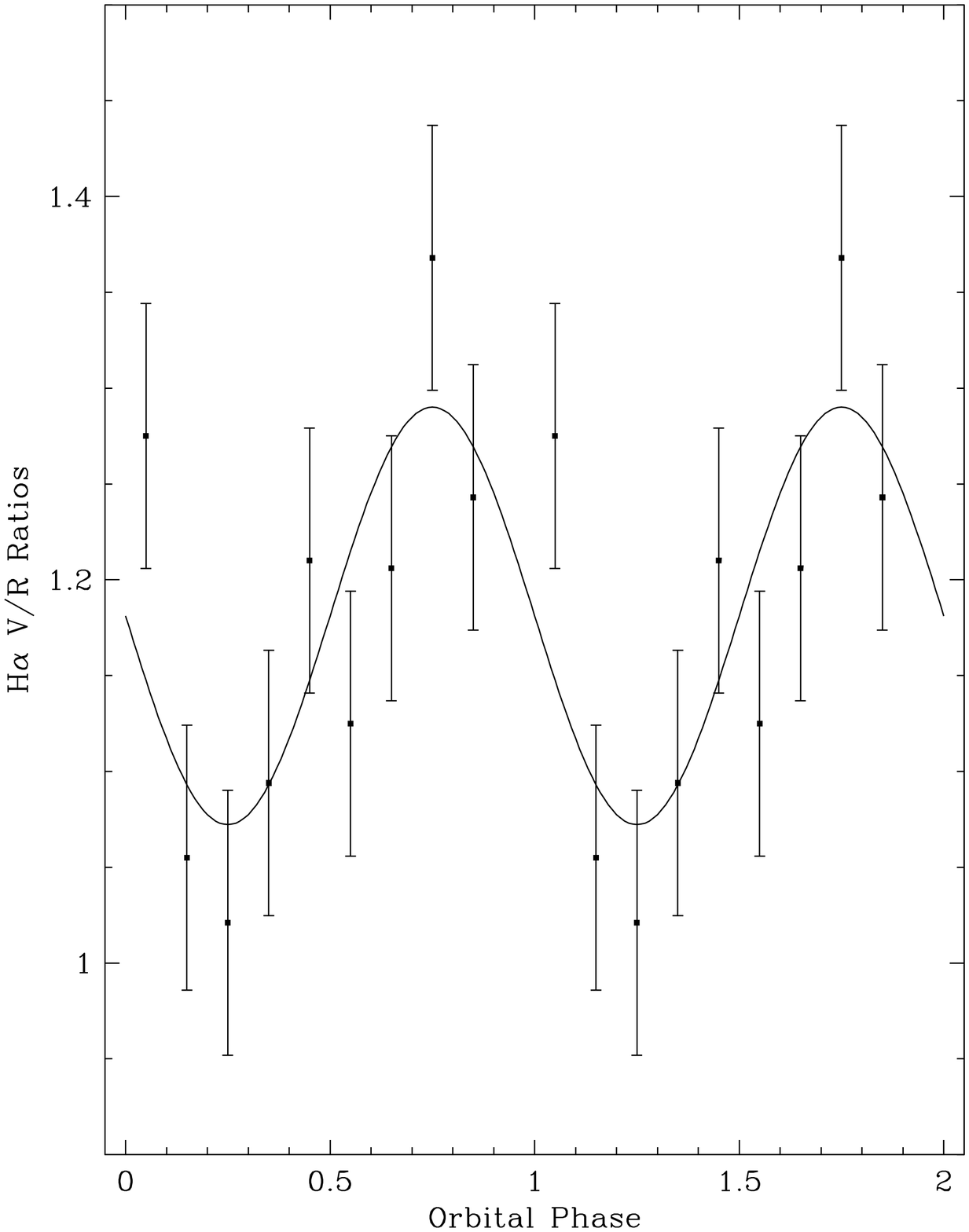}
     \caption{$Top$: Periodogram of V/R ratios for {\ha}.  $y$-axis is the normalised, reduced $\chi^2$.  Long dashed line shows 99\% confidence level level and dotted line the 68\% confidence level.  $Bottom$: Folded V/R ratios on the 14.86 hours orbital period.  The sine curve is best fit to data.  See text for fit parameters.}
     \label{fig:vr_red}
\end{centering}
\end{figure}

The periodogram of {\ha} V/R ratios (Figure \ref{fig:vr_red}) shows a clear peak at the published orbital period above the 99\% confidence level.  

The V/R ratios were folded on the orbital period using $T_o$ found in \S \ref{sec:indiv_rv} and binned by 0.10 in phase.  The results are shown in Figure \ref{fig:vr_red}.  There is an obvious modulation in the ratios.  

Sinusoid curves were fitted to the V/R modulation to obtain the semi-amplitude and mean value and their errors.  The errors on the binned ratios were taken from the standard deviation calculated from the least-squares sine fit.  

The minimum {\ha} V/R ratio is at orbital phase 0.25 and the maximum at phase 0.75.  This is consistent with the majority of flux variation originating from the accretion disk.  The semi-amplitude is 0.11 $\pm$ 0.04 and the ratios vary around a mean of 1.18 $\pm$ 0.03.  

\begin{figure}
\begin{centering}
\includegraphics[width=7cm,height=8cm]{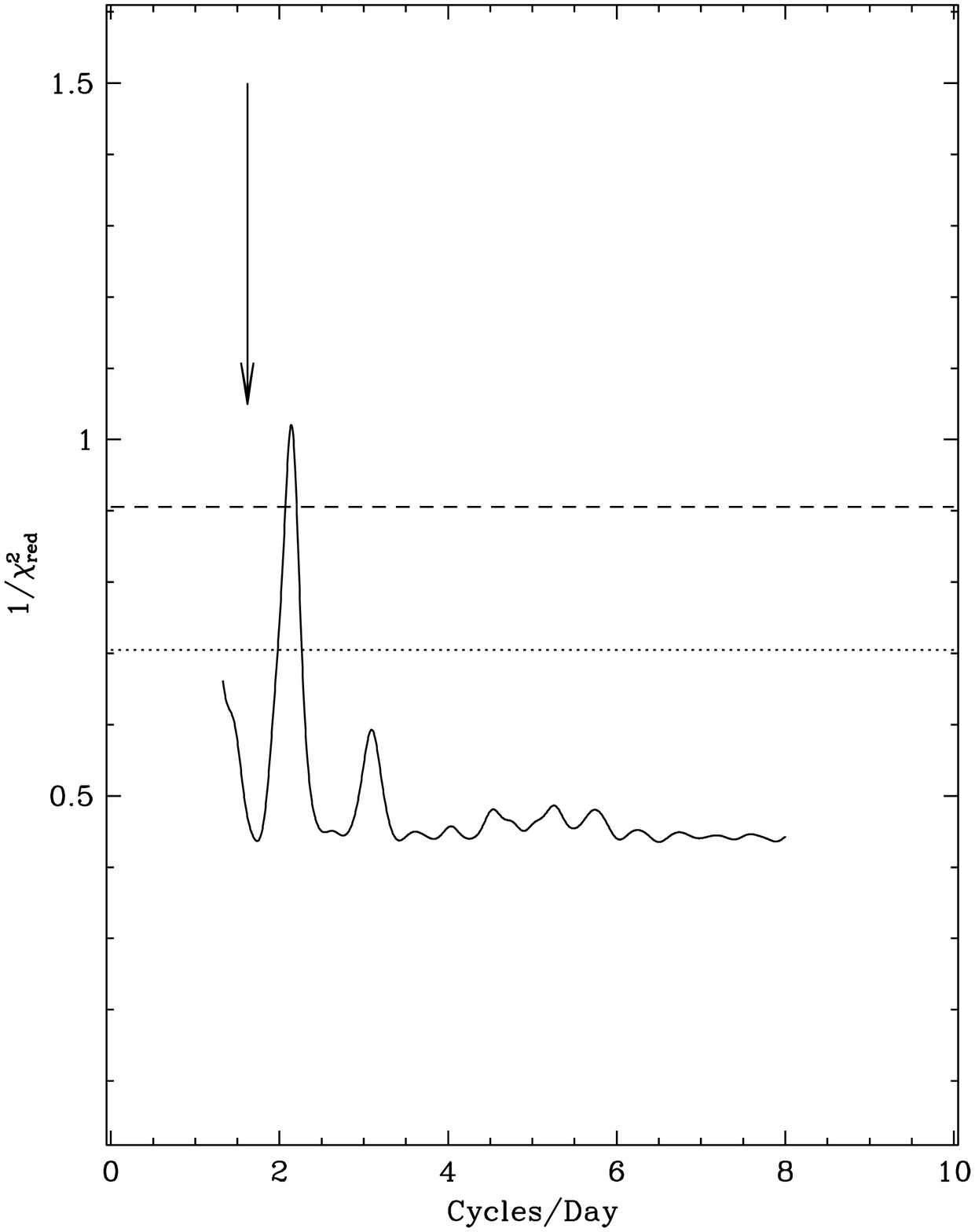}
\includegraphics[width=7cm,height=8cm]{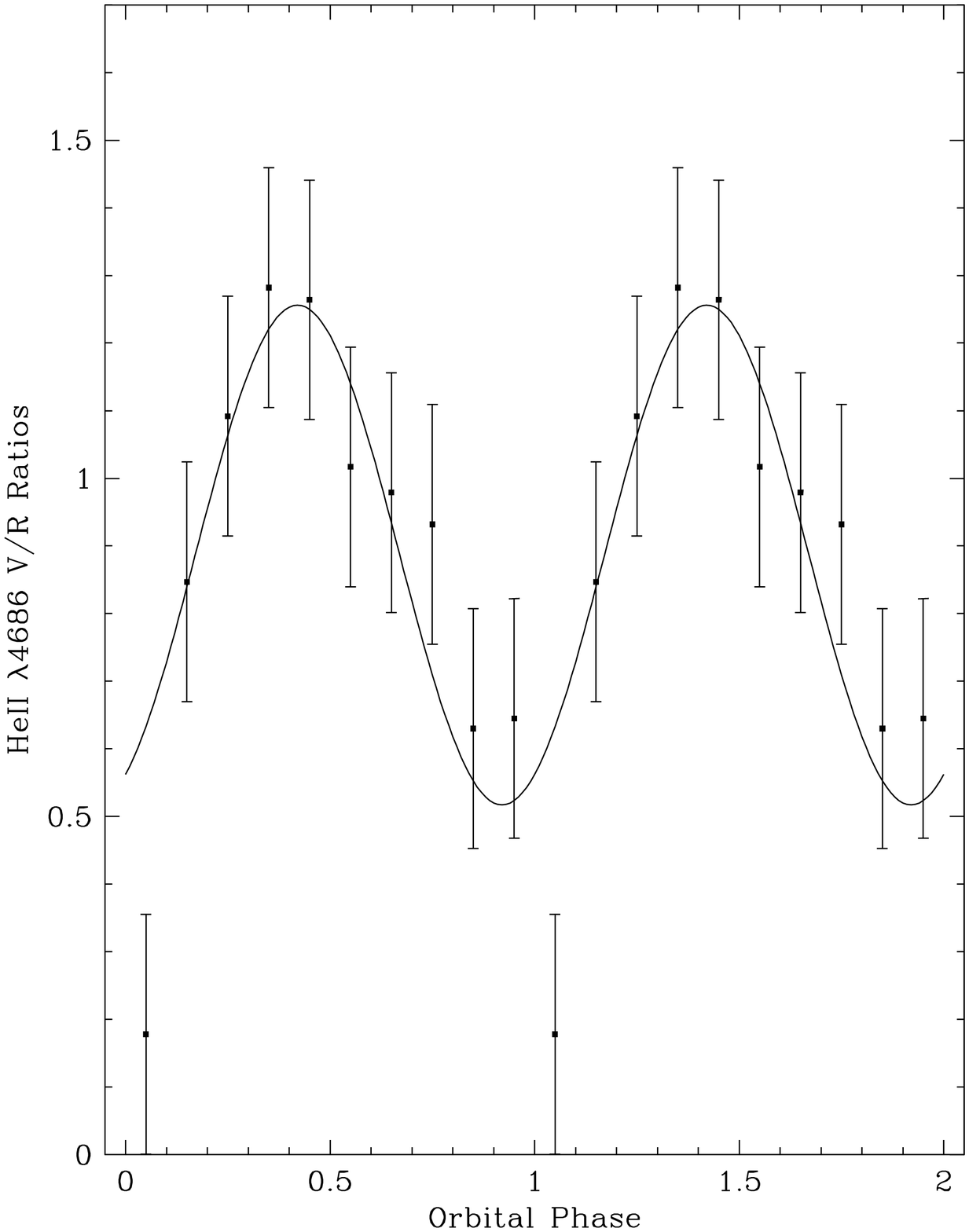}
     \caption{As for Figure \ref{fig:vr_red} for {\heii}.  V/R ratios folded on 11.2 hours corresponding to highest peak in periodogram.}
\label{fig:vr_blue}
\end{centering}
\end{figure}

V/R ratios were also measured for {\heii} and the resulting periodogram is shown in Figure \ref{fig:vr_blue}.  Although there is a single peak at the 99\% confidence level it does not correspond to published orbital period but rather $\sim$ 11.2 hours.  The V/R ratios of {\heii} were folded on this period and is presented in Figure \ref{fig:vr_blue}.  Besides one discrepant data point at phase 0.05, the data follow a well-defined sinusoid curve.  The sold curve is the best-fit sinusoid with a semi-amplitude of 0.37 $\pm$ 0.09 and mean 0.89 $\pm$ 0.06.  The maximum falls at phase $\sim$ 0.4 and minimum at phase $\sim$ 0.9.

\section{Discussion}

There is no doubt that measuring the orbital period of GX \nolinebreak339-4 via disk emission lines will always be difficult.  This is due to the low velocity amplitudes expected from a low-inclination \citep{wu01}.  This in itself demands good quality data, that is, data with high SNR and enough spectra to sample the orbital phases well.  It was clear in the analysis of spectra with low ($<$ 40) SNR and datasets with a low number of spectra that one could not obtain satisfactory periodograms.  This is why the results presented here have focussed on the dataset of 1998 May 28-31 as it had the best data quality.  

In the case of {\ha} the periodograms of the radial velocities, EWs and V/R ratios clearly show an isolated peak at the published orbital period at the 99\% confidence level.  

For {\heii} the radial velocity periodogram did not show a single peak but rather a broad double-peak around the published orbital period.  When one folds the radial velocities on 14.86 hours anyway one can see a clear deviation from an otherwise sinusoid curve at phases 0.75-0.95.  Whether this is due to some data defect or to a real physical source in the binary cannot be discerned at this point but it would be interesting to see if it is present in future observations.  What is clear is that the EW of {\heii} is modulated on the published orbital period and the minimum and maximum occur at the same phase as the {\ha} EW.  Since we expect {\ha} to originate from the outer disk regions and {\heii} further in, the source of the EW modulation must be the entire disk or at least a large part of it.  We leave this point here but return to this issue in Paper II when we study further the sources of the line profile variations.  The V/R ratios of {\heii} produced a periodogram with a single, isolated peak but \textit{not} at the published orbital period, rather at 11.2 hours.  It is not obvious why the {\heii} V/R ratios would modulate on a different period than {\ha}.  A clue as to why this may be the case is in the asymmetry of the line profiles.  {\ha} shows more asymmetry on the blue side than {\heii}.  The larger degree of asymmetry in {\ha} would translate into larger amplitudes of V/R ratios perhaps making it easier to discern an orbital modulation.  The V/R modulation of {\heii} is fairly clear, however.

By comparing the phases of minima and maxima of the radial velocity, EW and V/R ratio modulations it seems that the variations arise from the accretion disk.  In Paper II we attempt to answer whether these variations are due solely to a hotspot, to a disk which is asymmetric geometrically and/or in its emitting regions, or both.

\section{Summary and Conclusions}
\label{sec:gx_orbit_conclusion}

Optical spectra of GX 339-4 was obtained when the system was in a high (1998) and low (1999) X-ray state.

Using interstellar absorption lines from the spectra a distance of 4 $\pm$ 1 kpc was determined.  $E(B-V)$ was measured to be 1.1 $\pm$ 0.2.  These are consistent with previous studies.

Optical colours reddened from 1998 to 1999.  This is consistent with the accretion disk contributing less blue flux to the system due to the decrease in soft X-rays available to be reprocessed in the disk.

As the soft X-rays decreased in the low state the EW of {\ha}, {\hb} and {\heii} also decreased.  Without simultaneous photometry it is impossible to say whether this is mainly due to the continuum or line flux variations although the line flux must be declining faster than the continuum during the low state.  

The FWHM of the {\ha}, {\hb}, {\heii} and the {\bb} emission lines increased from 1998 to 1999.  This suggests that the line emitting regions moved closer to the compact object in 1999.  The inner regions of the accretion disk may have started to cool sufficiently for Balmer emission to occur there in the low state.  The higher ionisation lines of {\heii} and the {\bb} may only obtain sufficient photoionising X-rays closer to the compact object.  

The {\hb} emission line exhibited redshifted absorption at $\sim \lambda$4880 {\AA} during both the high and low X-ray states.  The origin of this absorption feature is not known.  

Emission line profiles vary on the timescales of hours.  Analysis of the 1998 May high resolution data show that the radial velocities, EWs and V/R ratios of {\ha} modulated on the published orbital period.  \textit{This is the first time that the published orbital period has been confirmed by spectroscopic data since the study of CCHT92.}  The EW of {\heii} also varied on the orbital period but the radial velocities and V/R ratios were inconclusive.  The radial velocities of the {\bb} modulated on the orbital period but peaked 0.25 out of phase with respect to {\ha}.  This result may not be reliable due to other significant peaks in the periodogram.  

The amplitude of the radial velocity modulation for {\ha} was 14 $\pm$ 3 {\kms} which is much less than that found by \citet{cal92}.  This, in turn, decreases the mass function.  Taking $P_{orb}$ = 14.86 hours, $M_2$ = 0.7 {\msun} and $i \le 70^o$ we find that $M_1 \le 40$ {\msun}.  If $i = 15^o$ \citep{wu01} then $M_1 = 5$ {\msun} which places the compact object in GX 339-4 into the black-hole category.  Determining the semi-amplitude velocity from emission lines may not be an accurate measurement of $K_{1}$ as has been well established from CV studies \citep{wad85}.  In order to establish whether our measurement of $K_{1}$ is reasonable future optical spectroscopic studies of GX 339-4 should endeavour to measure the semi-amplitude velocities of the emission lines using data of high resolution and SNR.  The semi-amplitude velocities of the same species (e.g. Balmer series) should be measured to see if they agree.  If not then we need to investigate the physical processes which cause such a disagreement which may lead to better methods of measuring an accurate $K_{1}$ from emission lines.  Until such time, the black-hole categorisation of GX 339-4 needs to be taken with caution.  

We may estimate the Roche lobe size ($R_L$) of the secondary star for $P_{orb}$ = 14.86 hours.  First, let us assume that the secondary is a K dwarf with mass $M_2$ = 0.7 {\msun}.  This is a reasonable assumption since the secondary stars in other low-mass X-ray binaries with similar periods are of this spectral type \citep[e.g. A0620-00,][]{mur80}.  If $M_1$ = 5 {\msun} then $R_L$ = 1.3 R$_{\odot}$.  If the secondary star in GX 339-4 is a K dwarf then it must be slightly evolved to fill its Roche lobe. If $M_1$ = 1.4 {\msun} then $R_L$ = 1.7 R$_{\odot}$.  Again, if the secondary is a K dwarf, it must be somewhat evolved.  

Future optical spectroscopic observations of GX 339-4 should endeavor to obtain simultaneous optical photometry to ascertain the cause of the EW modulation.  The absorption trough observed near {\hb} should be investigated further and it should be resolved whether {\ha} does or does not exhibit such a feature.  Most importantly, we must obtain as many spectra as possible with high SNR and good temporal coverage during various X-ray phases and at as many epochs as possible.  This would also enable us to ascertain whether the orbital modulation is observable only if certain physical conditions exist in the binary, for example, soft X-ray irradiation of a vertically extended hotspot.  

GX 339-4 is a difficult object to decipher due to the lack of information about the optical star and the complex behaviour of data.  We should, however, continue in our efforts to study GX 339-4.  This object transitions into all X-ray states more often than most and, therefore, provides many opportunities to learn about the outburst mechanisms and the structure of accretion disks.  

\section*{Acknowledgments}
The authors would like to thank Prof. Phil Charles and the referee for their useful comments.

\bsp

\label{lastpage}

\end{document}